\documentclass[12pt]{article}
\usepackage{setspace,graphicx,epstopdf,amsmath,amsfonts,amssymb,amsthm}
\usepackage{datetime}
\usepackage{float}
\usepackage[longnamesfirst]{natbib}
\usdate
\usepackage[english]{babel}
\usepackage{longtable}
\usepackage{ltablex,booktabs}
\usepackage{array,longtable,ragged2e,booktabs}
\newcolumntype{P}[1]{>{\RaggedRight\arraybackslash}p{#1}}
\usepackage{bbm}
\usepackage{array}
\usepackage{multirow}
\usepackage[tableposition=below]{caption}

\usepackage{tikz}
\usetikzlibrary{bayesnet}

\usepackage{dsfont}

\usepackage{booktabs}

\usepackage{multirow}
\usepackage{multicol}

\usepackage{enumerate}

\usepackage{subfig}

\makeatletter\let\chapter\@undefined\makeatother 

\usepackage{indentfirst} 
\usepackage{endnotes}    

\usepackage[labelfont=bf,labelsep=period]{caption}   

\usepackage[toc,page,titletoc]{appendix}

\newtheorem{definition}{Definition}[section]

\usepackage{chngcntr}
\counterwithin{figure}{section}

\makeatletter
\def\input@path{{Tables/}}
\makeatother

\begin{document}
	
	\doublespacing       
	
	\author{Juraj Hledik and Riccardo Rastelli\thanks{The authors are with the WU-Vienna University of Economics and Business. Welthandelsplatz 1, Vienna. The authors kindly acknowledge the financial support of the Austrian Science Fund (FWF) as well as the possibility to use data provided by Austrian National Bank (OeNB). 
	This research was also supported by the Vienna Science and Technology Fund (WWTF) Project MA14-031. All errors are our own responsibility.}} 
	
	\title{\Large \textbf{A dynamic network model to measure exposure diversification in the Austrian interbank market}}
	
	\date{\today} 
	
	
	\maketitle
	\thispagestyle{empty}
	\bigskip
	
	\centerline{\bf ABSTRACT}
	\begin{doublespace}
		We propose a statistical model for weighted temporal networks capable of measuring the level of heterogeneity in a financial system. Our model focuses on the level of diversification of financial institutions; that is, whether they are more inclined to distribute their assets equally among partners, or if they rather concentrate their commitment towards a limited number of institutions. Crucially, a Markov property is introduced to capture time dependencies and to make	our measures comparable across time. We apply the model on an original dataset of Austrian interbank exposures. The temporal span encompasses the onset and development of the financial crisis in 2008 as well as the beginnings of European sovereign debt crisis in 2011. Our analysis highlights an overall increasing trend for network homogeneity, whereby core banks have a tendency to distribute their market exposures more equally across their partners.
	\end{doublespace}
	\medskip
	
	\clearpage
	
	
	\newpage
\section*{Introduction}
During the past 10 years, the EU was hit by two major financial crises. In 2008, the problems started initially in the US subprime mortgage market and were partially caused by lax regulation and overly confident debt ratings. The source of the European sovereign debt crisis in 2011, however, was most likely private debt arising from property bubble and resulting in government bailouts. The lack of a common fiscal union in the EU did not help with the situation, which resulted in the European central bank providing cheap loans to maintain a steady cash flow between EU banks.
During these turbulent times, European banks were facing high levels of uncertainty. It was not clear which counterparty would remain solvent in the foreseeable future and even sovereign bonds were no longer considered the safest option. In the face of these unfavorable conditions, the banks were forced to reconsider their interbank investments and re-adjust their portfolios in order to account for the change in the economic situation.

In this paper, we study an original dataset of interbank exposures in Austria between the spring of 2008 and autumn of 2011. Namely, we introduce a dynamic network model to analyze banks exposures' diversification patterns as well as the overall trend towards diversification in the Austrian interbank market. To accomplish this task, we create an original latent variable model that allows one to analyze weighted networks evolving over time. This approach provides us with a model-based measure of systemic risk locally for each bank, but also globally for the financial system as a whole.
In our application, we show that our measure provides a qualitatively different view when compared to basic descriptive statistics. Tto achieve this, we resort to an intuitive modeling of a single network homogeneity (drift) parameter which we use to study the evolution of network homogeneity in time. Our model is specifically designed for instances where a network needs to be characterized by a single evolving variable, or when one is interested in obtaining a model-based quantitative measurement of the inter-temporal development of network homogeneity.

It is important to understand that a change in a financial network structure can have far-reaching and non-trivial consequences. To illustrate this fact further, consider a hypothetical financial network of four institutions (banks) represented by nodes and their mutual financial exposures (debt) represented by edges. In this simple example, connections are symmetric and every bank splits its investment among its neighbors equally. Furthermore, banks are required by a regulator to always keep a capital buffer to account for unexpected withdrawals, unfavorable economic conditions and other factors. Therefore, we assume that an institution remains safe unless it loses at least half of its investment. If that happens, the institution gets bankrupt and it might further negatively affect other banks in the network. To see how network structure affects the overall stability, consider a case where one of these four banks gets affected by an exogenous shock such that it has to declare bankruptcy. In such case, its neighbors will not get their respective investment and might suffer the same fate, putting their own neighbors in danger. This contagious behavior is dependent on how the banks are linked together, which illustrates the importance of structure when addressing questions on systemic importance and financial stability. 

\begin{figure}[ht!]
	\centering
	\subfloat[]{
		\label{fig:AnExample:DifferentStructures:a} 
		\includegraphics[width=0.2\linewidth]{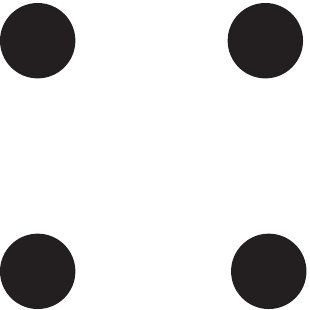}}
	\hspace{0.05\linewidth}
	\subfloat[]{
		\label{fig:AnExample:DifferentStructures:b} 
		\includegraphics[width=0.2\linewidth]{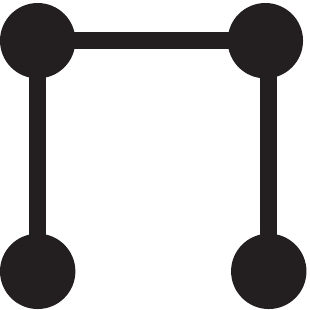}}
	\hspace{0.05\linewidth}
	\subfloat[]{
		\label{fig:AnExample:DifferentStructures:c} 
		\includegraphics[width=0.2\linewidth]{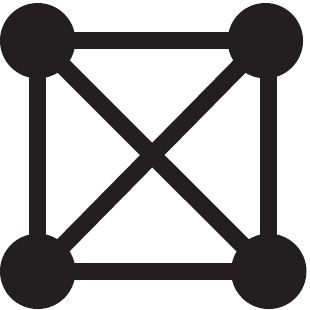}}
	\caption{Different loan network structures on a set of four banks.}
	\label{fig:AnExample:DifferentStructures} 
\end{figure}

For the hypothetical case of four banks, there are 11 different network structures that can possibly occur: a subset of these are shown in Figure \ref{fig:AnExample:DifferentStructures}.
In the case shown in Figure \ref{fig:AnExample:DifferentStructures:a}, there is no danger of contagion since there are no edges to propagate shocks. An analogical result follows from the network shown in Figure \ref{fig:AnExample:DifferentStructures:c}, where a failure of one node is not sufficient to take down the rest because every other institution only loses one third of its investment. Problems arise in intermediately connected systems such as \ref{fig:AnExample:DifferentStructures:b}, where an initial shock wipes out the whole system.

This basic example hints at a much more complex issue of network stability that has been extensively studied by financial regulators in the past two decades. More importantly, it highlights that the level of diversification in a system may play a crucial role in determining its stability and that assessment of this trait for observed networks can prove challenging. In this paper, we address this impasse, introducing a statistical model which is specifically designed to measure the diversification of a financial system, hence obtaining a measure for one of the facets of systemic risk.

This paper bridges two distinct academic fields. On the one hand, we strive to contribute towards the established literature on systemic risk and financial networks. This strand of literature has often focused on the stability of financial systems as well as the possibility of contagious bankruptcies similar to our simple example above. Research papers on this subject have been published by both academics in finance as well as market regulators.\footnote{This includes various country-specific central banks as well as the European Central Bank and the FED. Additional research has been undertaken by the Bank for International Settlements or the International Monetary Fund.} On the other hand, we also contribute towards theoretical papers dealing with latent variable modeling of network data. The method we propose borrows from and contributes to both fields, proposing a new perspective on systemic risk.

One of the earliest papers on the topic of systemic risk in finance was the work of \cite{Allen2000}, who have shown that the structure of the interbank market is important for the evaluation of possible contagious bankruptcies. Later on, \cite{Gai2010} extended their work from a simple model of four institutions to a financial network of an arbitrary size. Other notable papers on systemic risk include, for example, \cite{Glasserman2016} or \cite{acemoglu2015systemic}, while \cite{upper2011simulation} provides an excellent survey of regulatory-published scientific reports on the subject. With respect to the questions on diversification, we refer the reader to \cite{Elliott2014} and \cite{frey2014correlation} where a nontrivial relationship between diversification and contagious defaults is presented, or to \cite{goncharenko2015dark} where banks endogenously choose their level of diversification in an equilibrium setting. Our paper relates to these works, since it is the structure of a financial network we are studying, while exploring the questions regarding diversification at the same time. We further add to these papers introducing a new generative mechanism and a modelling framework where diversification and homogeneity of the system can be studied inter-temporally. 

As we have mentioned before, our paper also contributes to the research on latent variable modeling. Prominent examples include the latent position models of \cite{hoff2002latent}, later extended to the dynamic framework by \cite{sarkar2006dynamic}, and the latent stochastic blockmodels \citep{nowicki2001estimation} extended to a dynamic framework by \cite{yang2011detecting}, \cite{xu2014dynamic} and \cite{matias2017statistical}, among others. These latent variable models possess a number of desirable theoretical features, as illustrated in \cite{rastelli2016properties} and \cite{daudin2008mixture}, respectively.

Our approach also shares a number of similarities with other recent papers that apply a latent variable framework on various types of network data.
These include among others \cite{friel2016interlocking}, where the authors introduce a dynamic latent position model to measure the financial stability of the Irish Stock Exchange; 
\cite{sewell2016latent} who introduce a modeling framework for dynamic weighted networks;
but also \cite{mclaughlin2018empirical}, where the authors propose a framework to reconstruct a collaboration network.
Further related works include \cite{chakrabarti2017modeling}, where incentives of twitter users are analyzed; \cite{ji2016coauthorship} where meta-analysis of citations in statistics papers are conducted; 
and \cite{xin2017continuous}, where compatibility of basketball players are analyzed via a network model.
Lastly, we also contribute to the literature on the stability of the Austrian interbank market. Other works in this area include \cite{Elsinger2006}, \cite{Puhr2014} and \cite{Boss2004} who have looked at possible contagious effects and descriptive statistics of the Austrian financial network. We extend their work by creating a statistical model of network evolution. 

\section{Data and Exploratory Analysis}\label{sec:exploratory}
In this paper, we use a unique dataset obtained by the Austrian National Bank which contains quarterly observations of the Austrian Interbank Market for a period of four years (from spring of 2008 until autumn of 2011). More precisely, the dataset contains mutual claims between any two of $N=800$ Austrian banks during the corresponding quarter, resulting in 16 observations of the financial network. All of the banks considered existed throughout the whole period.

In order to comply with the privacy rules of the Austrian National Bank, the data is anonymized such that the true identities of banks in the system are hidden and replaced by non-descriptive IDs. Moreover, we are unable to see the true values of banks' mutual claims, only their scaled equivalents. Nevertheless, for the purposes of our model, the true values of connections in the financial network are not required, as we only need their relative size for a meaningful statistical analysis. In order to better clarify these concepts, we introduce the following terminology.

A \textit{dynamic network of interbank exposures} is a sequence of graphs where, for each time frame, the nodes correspond to banks and the edges correspond to the connections between them. In particular, the edges are directed and carry positive values indicating the claim of one bank to another. We note that an observed network of interbank exposures between $N$ banks over $T$ time frames may be represented as a collection of adjacency matrices of the same size $N\times N$, as in the following definition:

\begin{definition}
\label{def:TrueExposures}
A sequence of true exposures $\mathcal{E}=\{{E}_t\}_{t\in \mathcal{T}}$ defined on the set of nodes $\mathcal{V}$ over the timespan $\mathcal{T}$ consists of adjacency matrices $\textbf{E}_t \ \ \forall t \in \mathcal{T}$ with elements $e^{(t)}_{ij}$ for $t\in\mathcal{T}$, $i\in\mathcal{V}$, $j\in\mathcal{V}$, where $e^{(t)}_{ij}$ corresponds to the financial exposure of bank $i$ towards bank $j$ in period $t$.
\end{definition}

For the case of the Austrian interbank market, the adjacency matrix $\textbf{E}_t$ would contain the true values of all mutual claims between any two of $N=800$ Austrian banks at the corresponding time frame. However, as explained earlier, we are unable to observe the true exposures due to privacy policy of the Austrian National Bank. For the purpose of this paper, we will therefore be working with the sequence $\mathcal{Y}$ which corresponds to relative exposures from the creditor bank's point of view. More specifically, this transformation constricts the edge weights in our networks to a [0,1] interval, while keeping the sum of all exposures of a given bank equal to 1. For further details on the proper data transformation, description and derivation of $\mathcal{Y}$, please refer to the Appendix.

Additionally, we will be working with both the full dataset of 800 institutions (OeNB 800) as well as its subsample containing only the top 100 most heavily connected institutions which we will refer to as OeNB 100. Again, for further details on how we select the systemically relevant banks as well as the definition of bank relevance, refer to the Appendix.

\begin{figure}[!ht]
	\centering
	\begin{tabular}{cc}
		\subfloat[OeNB 800]{
			\label{fig:RelevanceEvolution:a} 
			\includegraphics[width=0.45\linewidth,page=3]{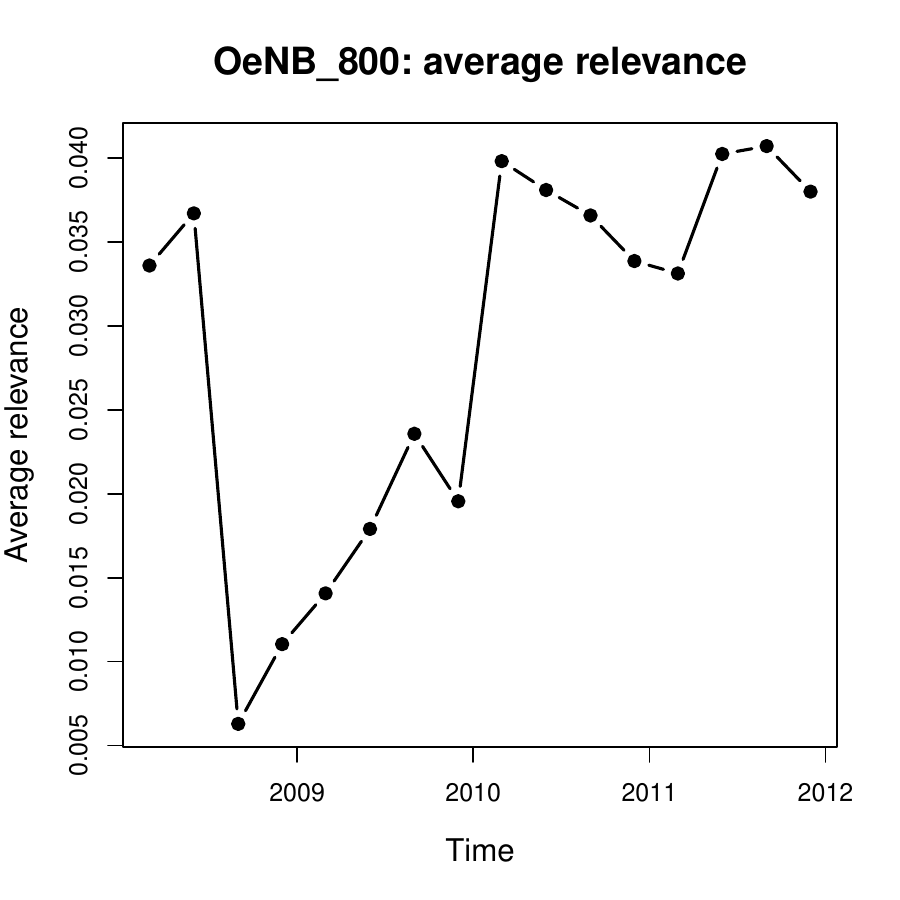}}
		\hspace{0.05\linewidth}
		&
		\subfloat[OeNB 100]{
			\label{fig:RelevanceEvolution:b}
			\includegraphics[width=0.45\linewidth,page=3]{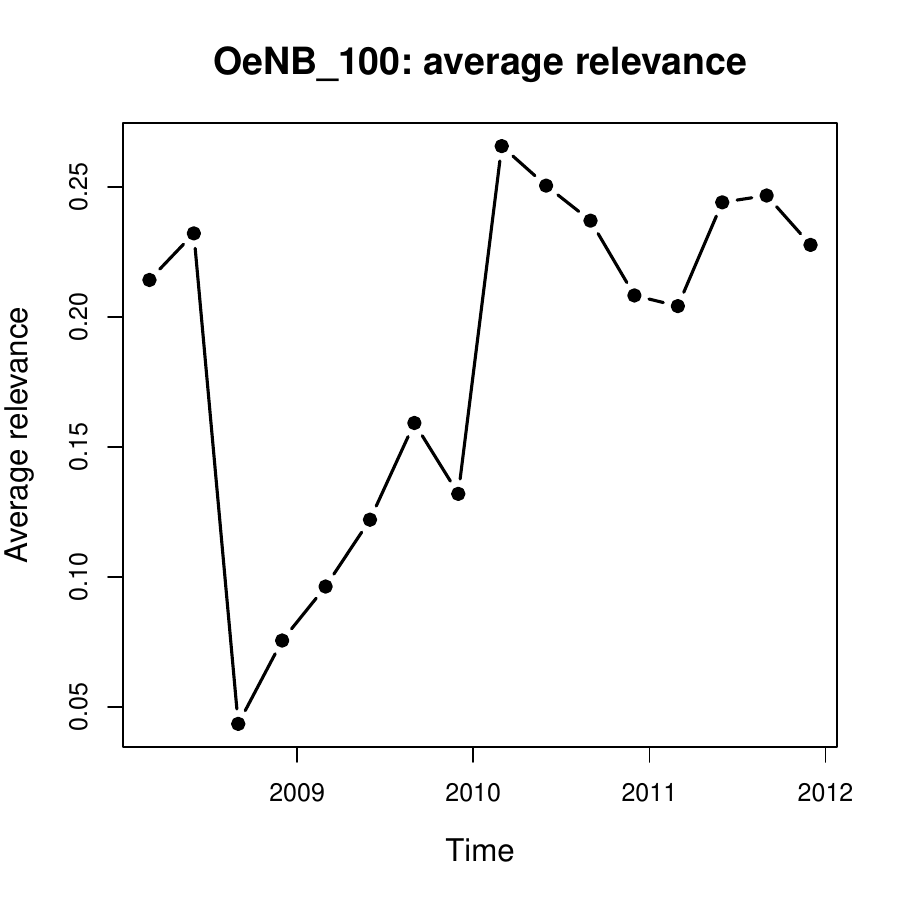}}
	\end{tabular}
	\caption{Bank relevance for the full sample (a) and the sample containing only the 100 most relevant banks (b).}
	\label{fig:RelevanceEvolution} 
\end{figure}

We plot the evolution of the average bank relevance in Figure \ref{fig:RelevanceEvolution}. Since the measure is directly proportional to the respective sizes of interbank exposures, we see a sharp drop in the second half of 2008 as a direct effect of the financial crisis. This drop corresponds to banks limiting the overall size of their exposures significantly in fear of counterparty risk.

\begin{figure}[!ht]
	\centering
	\includegraphics[width=0.45\linewidth]{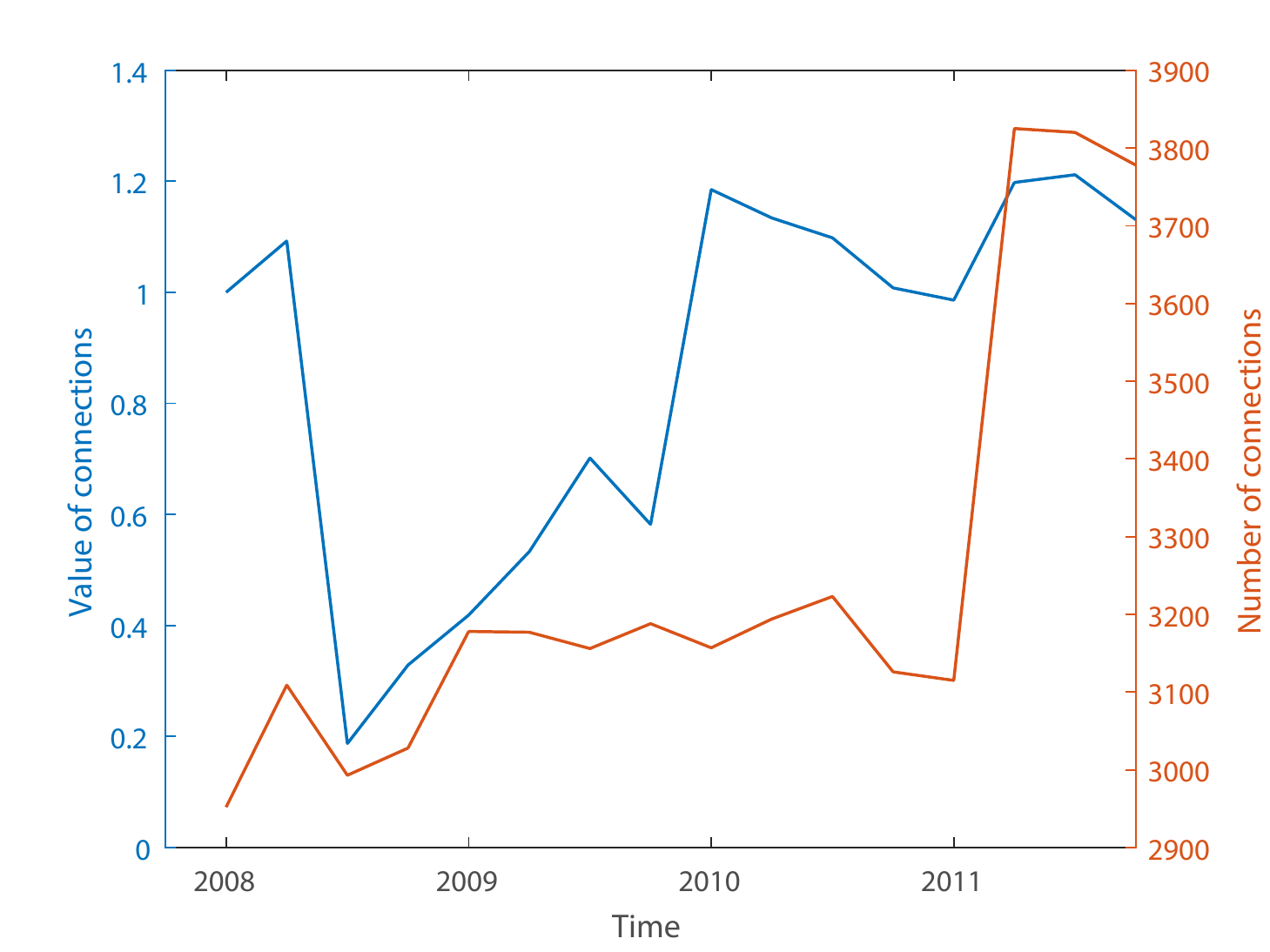}
	\caption{Number of connections and their relative size in time.}
	\label{fig:NumberAndValueOfEdges} 
\end{figure}

In order to have a better picture about the data, we have conducted a brief exploratory analysis of our dataset. In particular, it is interesting to see the evolution of connections in the sample. Table \ref{tab:DescriptiveStatistics} and Figure \ref{fig:NumberAndValueOfEdges} contain brief descriptive statistics, where one can see the number as well as magnitude of connections as a function of time. The number of connections ($2^{\text{nd}}$ column) shows the number of edges in the network as of time $t$, while the relative size ($3^{\text{rd}}$ column) depicts the overall cash flow in the market, scaled according to the first observation. We would like to highlight the second and third quarter of 2008, where a drop in the overall magnitude of cash flow in the economy can be observed. This period corresponds to the financial crisis associated with the failure of Lehman Brothers in the US and the problems stemming from the housing market. Interestingly, in the Austrian interbank market, the overall number of connections does not seem to be affected by these events as much as their size. This shows that, albeit Austrian banks have reduced their mutual exposures significantly, they were rarely completely cut off. Another important period is during the second and third quarter of 2011, which is roughly when the European sovereign debt crisis started. At the first glance, there does not seem to be much in relation to this event in our data. However, as we shall see later, our main model will provide further insight regarding the trend in diversification during this period.

\begin{figure}[!ht]
	\centering
	\begin{tabular}{cc}
		\subfloat[Adjacency matrix]{
			\label{fig:FirstPeriodMatrix:a} 
			\includegraphics[width=0.45\linewidth]{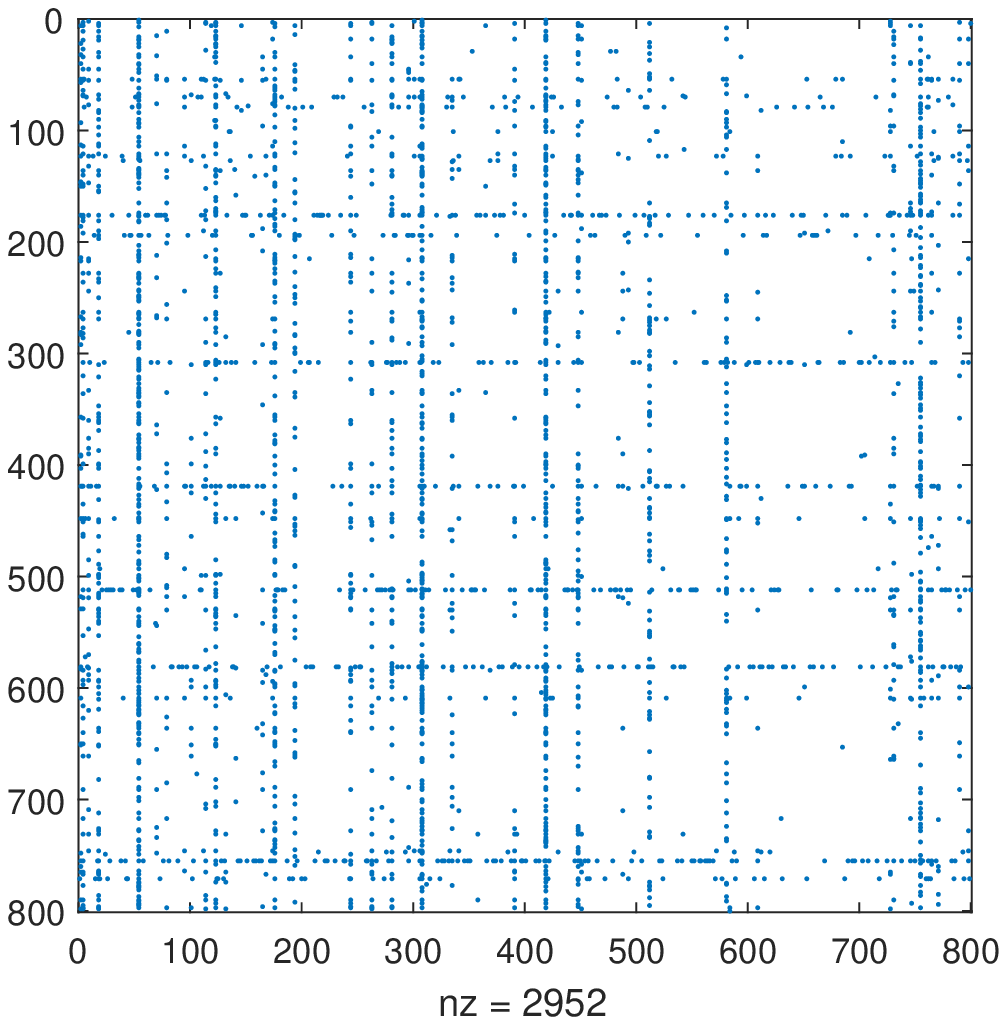}}
		\hspace{0.05\linewidth}
		&
		\subfloat[Plot of a network snapshot]{
			\label{fig:FirstPeriodMatrix:b}
			\includegraphics[width=0.40\linewidth]{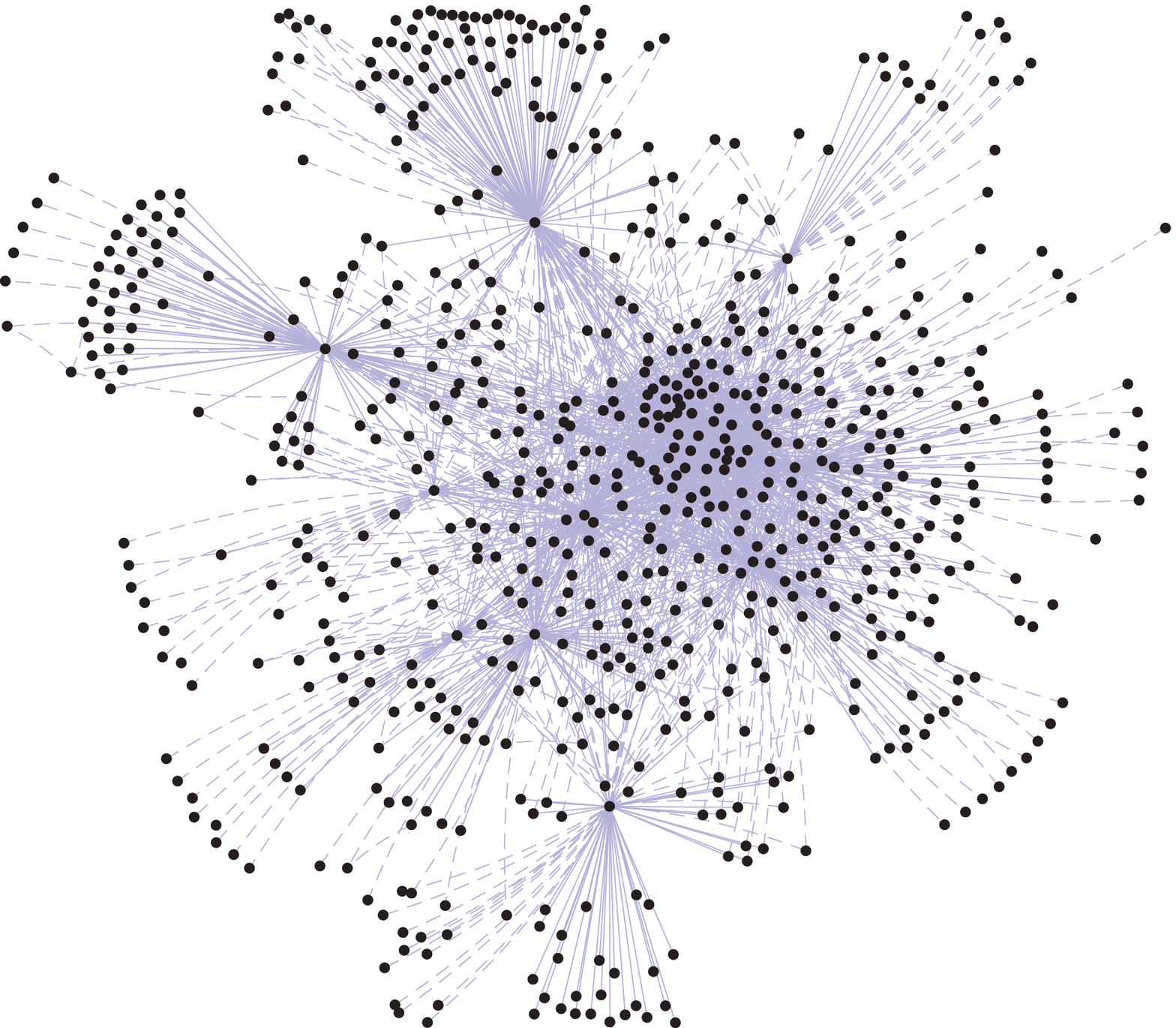}}
	\end{tabular}
	\caption{Adjacency matrix for the first time period, consisting of 2952 edges represented as dots (a) and a graphical representation of the network snapshot for the nodes with at least one connection (b).}
	\label{fig:FirstPeriodMatrix} 
\end{figure}

\begin{table}
\caption{Number of connections and their relative size in time.}
\label{tab:DescriptiveStatistics}
\centering
\begin{tabular}{ccc}
	   & No. of      & Relative size \\ 
Period & connections & of connections \\ 
\hline 
2008Q1 & 2952 & 1.0000 \\ 
2008Q2 & 3109 & 1.0925 \\ 
2008Q3 & 2993 & 0.1873 \\ 
2008Q4 & 3028 & 0.3287 \\ 
2009Q1 & 3178 & 0.4186 \\ 
2009Q2 & 3177 & 0.5329 \\ 
2009Q3 & 3156 & 0.7016 \\ 
2009Q4 & 3188 & 0.5820 \\ 
2010Q1 & 3157 & 1.1851 \\ 
2010Q2 & 3194 & 1.1340 \\ 
2010Q3 & 3223 & 1.0981 \\ 
2010Q4 & 3126 & 1.0080 \\ 
2011Q1 & 3115 & 0.9860 \\ 
2011Q2 & 3825 & 1.1979 \\ 
2011Q3 & 3820 & 1.2118 \\ 
2011Q4 & 3778 & 1.1310 \\
\hline  
\end{tabular} 
\end{table}
In interbank markets, it is common to observe disassortative properties in the system, which roughly translates to nodes with a low number of neighbors being connected to nodes with high number of neighbors and vice versa (see \cite{Hurd2016}) . This property in financial networks is quite common and different from social networks where individuals with a high number of ``friends'' tend to create ``hubs'' in the network, see for instance \cite{Li2014}. Financial systems also tend to be very sparse. We observe the same patterns in the Austrian interbank market, as can be seen from Figure \ref{fig:FirstPeriodMatrix}.

We have observed several interesting patterns in the data which suggest that using a more involved model could indeed produce some new insights to the evolution of bank diversification. Since the main interest of our research lies in the diversification of agents in an interbank market, we have also looked at the evolution of bank entropy in time. For this purpose, we use a standard definition of entropy as follows:
\begin{definition}
\label{def:NodeEntropy}
The \textit{entropy} $S_i^{(t)}$ of node $i\in \mathcal{V}$ at time $t \in \mathcal{T}$ is defined as:
\begin{equation}
S_i^{(t)}\stackrel{def}{=}-\sum_{k=1}^{N}y_{ik}^{(t)}\log{y_{ik}^{(t)}}
\end{equation}
\end{definition}
Speaking more plainly, this quantity describes how an institution distributes its assets among counterparties. A bank with only one debtor would have entropy equal to zero, since its relative exposure is one for that debtor and zero for all the other banks.
With an increased number of debtors with equal exposures, a node's entropy is increased and, for a fixed number of debtors, the entropy of a node is maximized when its assets are distributed evenly among neighbors. 
Ergo, if two nodes have the same number of outgoing connections, one may view the one with a higher entropy as better diversified. 
\begin{figure}[!ht]
	\centering
	\begin{tabular}{cc}
		\label{fig:Entropy:a} 
		\includegraphics[width=0.45\linewidth]{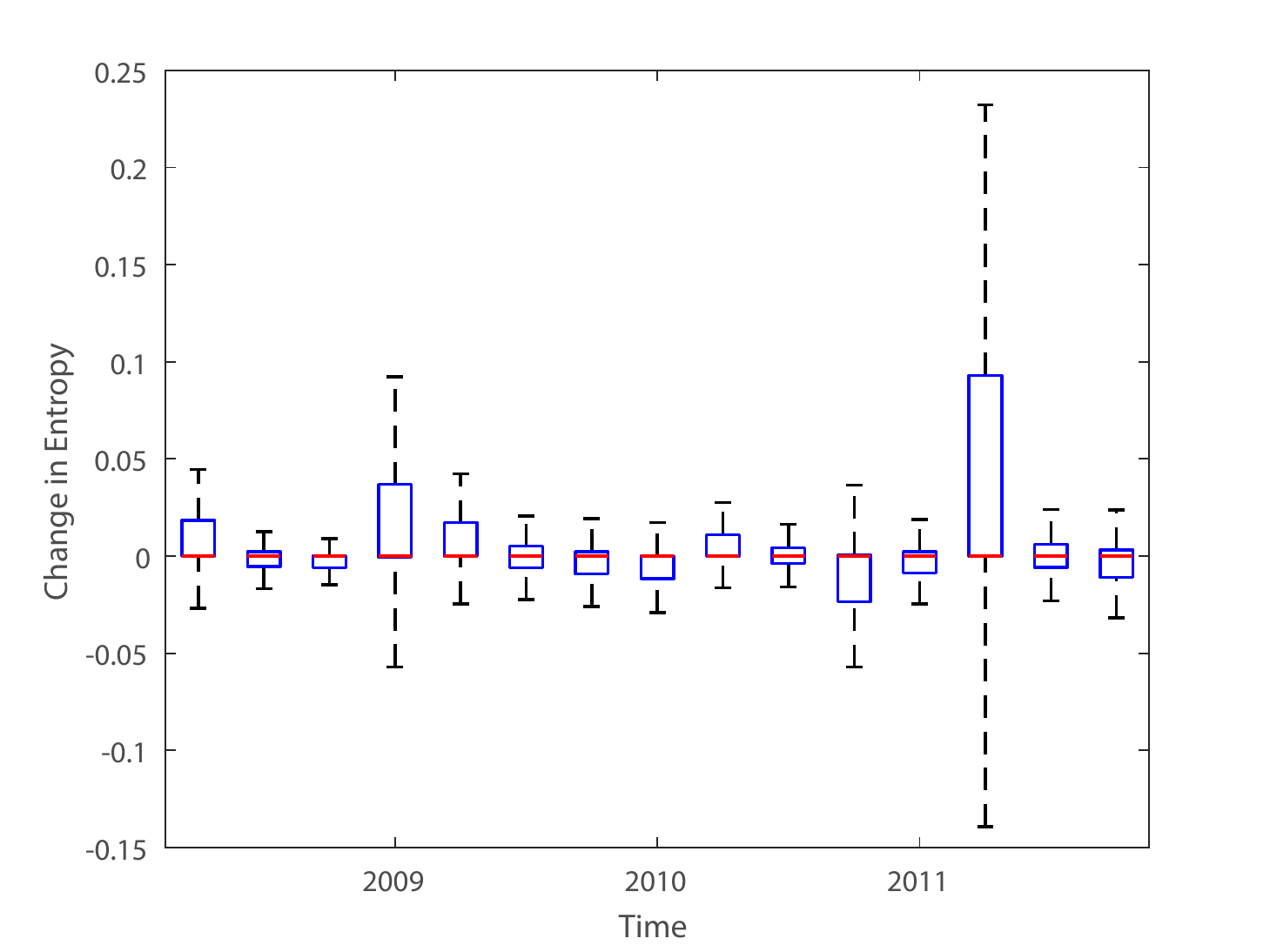}
	\end{tabular}
	\caption{Distribution of entropy change in time. }
	\label{fig:Entropy} 
\end{figure}

In Figure \ref{fig:Entropy}, we plot the change in nodes' entropies in consecutive periods ($S_i^{(t+1)} - S_i^{(t)}$). One can observe an increase in both mean and variance during the second and third quarter of 2011, which corresponds to the sovereign crisis in Europe. At that point, future bailouts of several EU countries were uncertain which might have added to the volatility in the market. Interestingly, no similar effect can be seen during the 2008 crisis.
In our model, we take all information from this exploratory analysis into account and we further focus on the diversification part of the story. In addition to the analysis on bank entropy, we extend this line of reasoning and create a measure of overall trend in diversification.
There are other ways of assessing the temporal evolution of node exposure homogeneity. We have chosen entropy for our exploratory analysis, as it constitutes a simple, clean and easily tractable approach, but one could easily turn to other measures, e.g. the Herfindahl index as is common practice in economics literature.

This data is observed quarterly, from the spring of 2008 to the autumn of 2012, hence, $T=16$ adjacency matrices are available in total (see Appendix \ref{app:data_transormation} for further details on data structure). The sequence of relative exposures is important in both the exploratory analysis we have conducted as well as in our main model. 

To model the dynamic evolution of a network, we assume discrete time steps in order to accommodate for our quarterly-observed data. A continuous time model in the spirit of \cite{Koskinen2012} would constitute a possible extension of our model.

Throughout the paper, we deal with different probability distributions. The normal distribution with mean $u$ and variance $v$ shall be denoted by $\mathcal{N}(u,v)$, the Gamma distribution with shape parameter $k$ and scale parameter $l$ shall be referred to as Gamma($k$,$l$) and a Dirichlet distribution parametrized by a vector \textbf{$\alpha$} shall be referred to as Dir($\boldsymbol{\alpha}$).\footnote{Notable papers on Dirichlet distribution and its usage include \cite{Minka2000}, \cite{Merwe2018}, and \cite{hijazi_modelling_2009}.
}

\section{The Model} \label{sec:TheModel}
We use the relative interbank exposures $y_{ij}^{(t)}$ from Definition \ref{def:RelativeExposures}, assuming that there are no self-connections such that when not stated otherwise, we always work with $t\in\mathcal{T}$, $i,j \in \mathcal{V}$ and $i\neq j$.
As these are relative exposures, it follows from definition that they satisfy:
\begin{equation}
 y_{ij}^{(t)} \in [0,1] \hspace{1cm}\mbox{and}\hspace{1cm} \sum_{j\in\mathcal{V}: j\neq i}y_{ij}^{(t)} = 1.
\end{equation}

We propose to model the vector $\textbf{y}_{i\cdot}^{(t)} = \left( y_{i1}^{(t)}, \dots, y_{iN}^{(t)} \right)$ as a Dirichlet random vector characterized by the parameters 
$\boldsymbol{\alpha}_{i\cdot}^{(t)} = \left( \alpha_{i1}^{(t)}, \dots, \alpha_{iN}^{(t)} \right)$, where $\alpha_{ij}^{(t)}>0$.
Following the established standard in latent variable models, the data are assumed to be conditionally independent given the latent parameters $\boldsymbol{\alpha} = \left\{\alpha_{ij}^{(t)}\right\}_{i,j,t}$.
Hence, the model likelihood reads as follows:
\begin{equation}
\mathcal{L}_{\mathcal{Y}}\left( \boldsymbol{\alpha} \right) = \prod_{t=1}^{T}\prod_{i=1}^{N}\left\{\frac{\Gamma\left( \sum_{j}y_{ij}^{(t)} \right)}{\prod_{j}\Gamma\left( y_{ij}^{(t)} \right)} \prod_j \left[y_{ij}^{(t)}\right]^{\alpha_{ij}^{(t)}-1}  \right\}
\end{equation}
where, again, $j$ varies in $\mathcal{V}$ and is different from $i$, and $\Gamma\left( \cdot \right)$ denotes the gamma function.

As concerns the $\boldsymbol{\alpha}$ parameters, we separate a trend component from the sender and receiver random effects through the following deterministic representation:
\begin{equation}\label{eq:alpha_1}
\log\left(\alpha_{ij}^{(t)}\right) = \mu_t + \theta_i + \gamma_j.
\end{equation}
With this formulation, the model parameters $\boldsymbol{\mu} = \left\{\mu_t\right\}_{t\in\mathcal{T}}$, $\boldsymbol{\theta} = \left\{\theta_i\right\}_{i\in\mathcal{V}}$ and $\boldsymbol{\gamma} = \left\{\gamma_j\right\}_{j\in\mathcal{V}}$ possess a straightforward interpretation.

\subsection{Interpretation of model parameters}
Before we move to parameter interpretation, we would like to note that the effect of $\alpha$ on a symmetric random vector $\textbf{Y}\sim Dir\left( \alpha,\dots,\alpha \right)$. Namely, it is important to see that the variance of $\textbf{Y}$ decreases with an increase in $\alpha$.
Since the values generated from a Dirichlet distribution lie in an $(N-1)$-dimensional simplex, low variance translates to $y_i \approx 1/\left( N-1 \right), \forall i \in \mathcal{V}$, e.g. the values are more or less equally distributed. High variance, however, implies that one of the components turns out to be close to one while all the others are close to zero.
This mechanic closely mimics the high-entropy homogeneous regime and the low-entropy heterogeneous regime introduced in Section \ref{sec:exploratory}, respectively.

In fact, in our formulation, the contribution given by $\mu_t + \theta_i$ affects all of the components of $\boldsymbol{\alpha}_{i\cdot}^{(t)}$ in a symmetric fashion.
Hence, we are essentially capturing the level of homogeneity in the network through a homogeneity trend parameter $\mu_t$ and a node specific homogeneity random effect $\theta_i$.
In other words, an increase in $\mu_t + \theta_i$ corresponds to higher diversification of exposures for bank $i$ at time $t$, resulting in a more homogeneous network structure.
Vice versa, a decrease in $\mu_t + \theta_i$ is linked with a decrease in diversification which in turn results in a more heterogeneous network structure.

The interpretation of $\gamma_j$ is similar. 
To see this, consider a non-symmetric random vector $\textbf{Y}\sim Dir\left( \alpha_1,\dots,\alpha_N \right)$. 
In this case, an increase in a single parameter component $\alpha_j$ determines a higher expected value in $y_j$, at the expense of the other elements in $\textbf{Y}$.
In our context, an increase in $\gamma_j$ tends to increase the weight of all edges that $j$ receives from its counterparties. Equivalently, one can say that in such case the bank $j$ becomes more attractive, in the spirit of other banks concentrating their exposures more towards $j$.

To summarize, there is a clear way to interpret the main parameters of our model. Parameter $\mu_t$ indicates the global homogeneity level at time frame $t\in\mathcal{T}$, parameter $\theta_i$ characterizes the individual bank $i$ homogeneity level as a random effect, and parameter $\gamma_j$ represents the bank $j$'s attractiveness.

\subsection{Bayesian hierarchical structure}
We complete our model by introducing the following Bayesian hierarchical structure on the parameters we have mentioned earlier.

We assume a random walk process prior on the drift parameters $\boldsymbol{\mu}$ as follows:
\begin{align*}
\mu_1 \sim \mathcal{N}(0,1/\tau_{\mu}),& &\mu_t=\mu_{t-1} + \eta_t, \ \ \forall t>1,
\end{align*}
where $\eta_t\sim \mathcal{N}(0,1/\tau_\eta)$ and $\tau_\eta \sim Gamma(a_\eta,b_\eta)$.
The hyperparameter $\tau_{\mu}$ is user-defined and set to a small value to support a wide range of initial conditions.
The hyperparameters $a_\eta$ and $b_\eta$ are also user-defined and set to small values ($0.01$) to allow a flexible prior structure.

The parameters $\boldsymbol{\theta}$ and $\boldsymbol{\gamma}$ are assumed to be i.i.d. Gaussian variables with:
\begin{align*}
\theta_i \sim \mathcal{N}\left(0,\frac{1}{\tau_\theta}\right),& &\tau_\theta \sim \text{Gamma}(a_\theta,b_\theta),\\
\gamma_j \sim \mathcal{N}\left(0,\frac{1}{\tau_\gamma}\right),& &\tau_\gamma \sim \text{Gamma}(a_\gamma,b_\gamma)
\end{align*}
Similarly to the other hyperparameters, $a_\theta$, $b_\theta$, $a_\gamma$ and $b_\gamma$ are also set to small values ($0.01$).

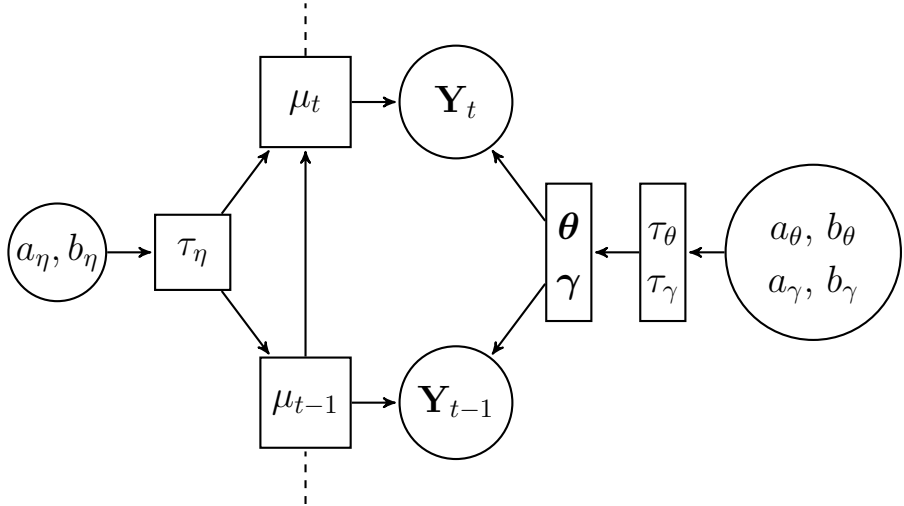
\begin{figure}[!ht]
\begin{center}
\begin{tikzpicture}[->,>=stealth',scale=1,shorten >=1pt,auto,node distance=3cm,thick,main node/.style={circle,draw,font=\sffamily\large\bfseries}]
  \node[obs, shape=circle, xshift=1cm, fill=white, minimum size=1cm]  (hypertaumu) at (-0.3,0) {\large{$a_\eta,b_\eta$}};
  \node[obs, shape=rectangle, xshift=1cm, fill=white, minimum size=1cm]  (taumu) at (1.5,0) {\large{$\tau_\eta$}};
  \node[obs, shape=rectangle, xshift=1cm, fill=white, minimum size=1.2cm]  (mutm1) at (3,-2) {\large{$\mu_{t-1}$}};
  \node[obs, shape=rectangle, xshift=1cm, fill=white, minimum size=1.2cm]  (mut) at (3,2) {\large{$\mu_{t}$}};
  \node[obs, shape=circle, xshift=1cm, fill=white, minimum size=1.5cm]  (Ytm1) at (5,-2) {\large{$\textbf{Y}_{t-1}$}};
  \node[obs, shape=circle, xshift=1cm, fill=white, minimum size=1.5cm]  (Yt) at (5,2) {\large{$\textbf{Y}_{t}$}};
  \node[obs, shape=rectangle, xshift=1cm, fill=white, minimum size=0.6cm, align=center]  (thetagamma) at (6.5,0) {\\ \\ \large{$\boldsymbol{\theta}$}\\ \\ \large{$\boldsymbol{\gamma}$}\\ };
  \node[obs, shape=rectangle, xshift=1cm, fill=white, minimum size=0.6cm, align=center]  (tauthetataugamma) at (7.75,0) {\\ \\ \large{$\tau_{\theta}$}\\ \\ \large{$\tau_{\gamma}$}\\ };
  \node[obs, shape=circle, xshift=1cm, fill=white, minimum size=1cm, inner sep = 0, align=center]  (hypertauthetataugamma) at (9.75,0) {\\ \\ \large{ $a_\theta$,$\ b_\theta\ $}\\ \\ \large{ $a_\gamma$,$\ b_\gamma\ $}\\ };
  \node[] (invtopmu) at (4,3.5) {};
  \node[] (invbotmu) at (4,-3.5) {};
  \draw (hypertaumu) -- (taumu) node [midway, fill=white, scale=0.5, above=-0.35] {};
  \draw (taumu) -- (mutm1) node [midway, fill=white, scale=0.5, above=-0.35] {};
  \draw (taumu) -- (mut) node [midway, fill=white, scale=0.5, above=-0.35] {};
  \draw (mutm1) -- (mut) node [midway, fill=white, scale=0, above=-0.35] {};
  \draw (mutm1) -- (Ytm1) node [midway, fill=white, scale=0, above=-0.35] {};
  \draw (mut) -- (Yt) node [midway, fill=white, scale=0, above=-0.35] {};
  \draw (thetagamma) -- (Ytm1) node [midway, fill=white, scale=0.5, above=-0.35] {};
  \draw (thetagamma) -- (Yt) node [midway, fill=white, scale=0.5, above=-0.35] {};
  \draw (tauthetataugamma) -- (thetagamma) node [midway, fill=white, scale=0.5, above=-0.35] {};
  \draw (hypertauthetataugamma) -- (tauthetataugamma) node [midway, fill=white, scale=0.5, above=-0.35] {};
  \draw [-,dashed] (mut) to (invtopmu);
  \draw [-,dashed] (invbotmu) to (mutm1);
\end{tikzpicture}
\end{center}
\caption{Graphical representation of model dependencies.}
\label{fig:graphical_model} 
\end{figure}

The arrangement of parameters in Figure \ref{fig:graphical_model} summarizes the dependencies in our model graphically.

\section{Parameter estimation}
Our proposed model has $T$ drift parameters ($\boldsymbol{\mu}$), $N$ diversification parameters ($\boldsymbol{\theta}$), $N$ attractiveness parameters ($\boldsymbol{\gamma}$), and $3$ precision parameters ($\boldsymbol{\tau}$). We use this section to describe their estimation procedure.

\subsection{Identifiability}
The additive structure in \eqref{eq:alpha_1} yields a non-identifiable likelihood model. 
For example, one may define $\tilde{\theta}_i = \theta_i + c$ and $\tilde{\gamma}_j = \gamma_j - c$ for some $c\in\mathbb{R}$ and the likelihood value would be the same for the two configurations, i.e. $\mathcal{L}_{\mathcal{Y}}\left( \mu,\tilde{\theta},\tilde{\gamma} \right) = \mathcal{L}_{\mathcal{Y}}\left( \mu,\theta,\gamma \right)$.
One way to deal with such identifiability problem would be to include a penalization through the priors on $\boldsymbol{\theta}$ and $\boldsymbol{\gamma}$. One could specify more informative Gaussian priors centered in zero, which would in turn shrink the parameters to be distributed around zero.

However, such approach may also interfere with the results, since the model would not be able to capture the presence of outliers. Hence, we opt for a more commonly accepted method, and impose the $\boldsymbol{\gamma}$s to sum to zero.
This is expressed through the following constraint:
\begin{equation}
 \gamma_1 = -\sum_{j=2}^N\gamma_j.
\end{equation}

This new model, characterized by $T+2N+2$ parameters, is now identifiable.

\subsection{Markov chain Monte Carlo}
The posterior distribution associated to our model factorizes as follows:
\begin{equation}\label{eq:posterior_1}
\begin{split}
 &\hspace{0.1cm}\pi\left( \boldsymbol{\mu},\boldsymbol{\theta},\boldsymbol{\gamma},\tau_{\eta},\tau_{\theta},\tau_{\gamma} \right) \propto \\
 &\hspace{1cm}\propto\mathcal{L}_{\mathcal{Y}}\left( \boldsymbol{\mu},\boldsymbol{\theta},\boldsymbol{\gamma} \right)
 \hspace{0.1cm}\pi\left( \boldsymbol{\mu}\middle\vert \tau_{\eta}\right)\pi\left( \tau_{\eta}\middle\vert a_\eta,b_\eta \right)
 \hspace{0.1cm}\pi\left( \boldsymbol{\theta}\middle\vert \tau_{\theta}\right)\pi\left( \tau_{\theta}\middle\vert a_\theta,b_\theta \right)
 \hspace{0.1cm}\pi\left( \boldsymbol{\gamma}\middle\vert \tau_{\gamma}\right)\pi\left( \tau_{\gamma}\middle\vert a_\gamma,b_\gamma \right)
\end{split}
\end{equation}

We adopt a fully Bayesian approach, relying on a Markov chain Monte Carlo to obtain a random sample from the posterior distribution \eqref{eq:posterior_1}.
Note that, in the following equations, the products are defined over the spaces $\mathcal{T}$ and $\mathcal{V}$, with the only restriction that $j$ and $\ell$ are always different from $i$.
Also, $\mathds{1}_{\mathcal{A}}$ is equal to $1$ if the event $\mathcal{A}$ is true or zero otherwise.
We use a Metropolis-within-Gibbs sampler that alternates the following steps:
\begin{enumerate}
  \item Sample $\mu_s$ for all $s\in\mathcal{T}$ from the following full-conditional using Metropolis-Hastings with a Gaussian proposal:
  \begin{equation}
  \begin{split}
   \pi\left( \mu_{s}\middle\vert \dots \right) &\propto
   \left\{\prod_{i} \Gamma\left( e^{\mu_{s}}e^{\theta_i}\sum_je^{\gamma_j} \right)\right\}
   \left\{\prod_{i,j} \frac{  \left[y_{ij}^{(s)}\right]^{\alpha_{ij}^{(s)}-1}   }{   \Gamma\left(  \alpha_{ij}^{(s)}\right)   }\right\} \\
   &\hspace{1cm}\cdot \left\{ \exp\left\{-\frac{\tau_{\mu}\left[\mu_{s}\right]^2}{2}\right\} \right\}^{\mathds{1}_{\left\{s=1\right\}}} \\
   &\hspace{1cm}\cdot \left\{ \exp\left\{-\frac{\tau_{\eta}\left[\mu_{s}-\mu_{s-1}\right]^2}{2}\right\} \right\}^{\mathds{1}_{\left\{s>1\right\}}} \\
   &\hspace{1cm}\cdot \left\{ \exp\left\{-\frac{\tau_{\eta}\left[\mu_{s+1}-\mu_{s}\right]^2}{2}\right\} \right\}^{\mathds{1}_{\left\{s<T\right\}}}.
  \end{split}
  \end{equation}
  \item Sample $\theta_k$ for all $k\in\mathcal{V}$ from the following full-conditional using Metropolis-Hastings with a Gaussian proposal:
  \begin{equation}
   \pi\left( \theta_k\middle\vert \dots \right) \propto
   \left\{\prod_{t} \Gamma\left( e^{\mu_{t}}e^{\theta_k}\sum_je^{\gamma_j} \right)\right\}
   \left\{\prod_{t,j} \frac{  \left[y_{kj}^{(t)}\right]^{\alpha_{kj}^{(t)}-1}   }{   \Gamma\left( \alpha_{kj}^{(t)} \right)  }\right\} \exp\left\{-\frac{\tau_{\theta}}{2}\theta_k^2\right\}.
  \end{equation}
  \item Sample $\gamma_\ell$ for all $\ell\in\mathcal{V}\setminus\{1\}$ from the following full-conditional using Metropolis-Hastings with a Gaussian proposal:
  \begin{equation}
  \begin{split}
   \pi\left( \gamma_{\ell}\middle\vert \dots \right) &\propto
   \left\{\prod_{t,i} \Gamma\left( e^{\mu_{t}}e^{\theta_i}\sum_je^{\gamma_j} \right)\right\}
   \left\{\prod_{t,i} \frac{  \left[y_{i\ell}^{(t)}\right]^{\alpha_{i\ell}^{(t)}-1}   }{   \Gamma\left( \alpha_{i\ell}^{(t)} \right)  }\right\} \\
   &\hspace{1cm}\cdot\left\{\prod_{t,i} \frac{  \left[y_{i1}^{(t)}\right]^{\alpha_{i1}^{(t)}-1}   }{   \Gamma\left( \alpha_{i1}^{(t)} \right)  }\right\}
   \exp\left\{-\frac{\tau_{\gamma}}{2}\gamma_{\ell}^2\right\}.
  \end{split}
  \end{equation}
  \item Sample $\tau_{\eta}$ from the following conjugate full-conditional:
  \begin{equation}
   \pi\left( \tau_{\eta} \middle\vert\dots \right) \sim Gamma\left( a_{\eta} + \frac{T-1}{2}  ,  b_{\eta} + \sum_{t>1}\left( \mu_{t}-\mu_{t-1} \right)^2/2 \right).
  \end{equation}
  \item Sample $\tau_{\theta}$ from the following conjugate full-conditional:
  \begin{equation}
   \pi\left( \tau_{\theta} \middle\vert\dots \right) \sim Gamma\left( a_{\theta} + N/2  ,  b_{\theta} + \sum_{i}\theta_i^2/2 \right).
  \end{equation}
  \item Sample $\tau_{\gamma}$ from the following conjugate full-conditional:
  \begin{equation}
   \pi\left( \tau_{\gamma} \middle\vert\dots \right) \sim Gamma\left( a_{\gamma} + \frac{N-1}{2}  ,  b_{\gamma} + \sum_{j>1}\gamma_j^2/2 \right).
  \end{equation}
\end{enumerate}

The random draws obtained for the model parameters are then used to characterize their posterior distribution given the data.

\subsection{Empirical Analysis}
We ran our Metropolis-within-Gibbs sampler on both datasets \texttt{OeNB 800} and \texttt{OeNB 100} for a total of $400{,}000$ iterations.
For both datasets, the first $200{,}000$ iterations were discarded as burn-in. 
For the remaining sample, every $20$-th draw was saved to produce the final results.
In summary, we obtained $10{,}000$ posterior draws for each model parameter.

The first $100{,}000$ iterations of the burn-in period were also used to adaptively tune the Gaussian proposal variance individually for each parameter, to make sure that all of the acceptance rates were between $22\%$ and $30\%$. The variances were hence fixed to the these values thereafter the trace plots and convergence diagnostic tests all showed very good mixing of the Markov chain, suggesting a satisfactory convergence.

Similarly to many other latent variable models for networks, the computational cost required by our sampler grows as $TN^2$.
We implemented the algorithm in \texttt{C++} and used parallel computing via the library \texttt{OpenMPI} to speed up the procedure.
We note that, for the full dataset, an iteration required an average of approximately $0.75$ seconds on a Debian machine with $16$ cores.
The code is available from the authors upon request.

\section{Results}
\begin{figure}[t]
	\centering
	\begin{tabular}{cc}
		\subfloat[OeNB 800]{
			\label{fig:AvgMu:a} 
			\includegraphics[width=0.45\linewidth]{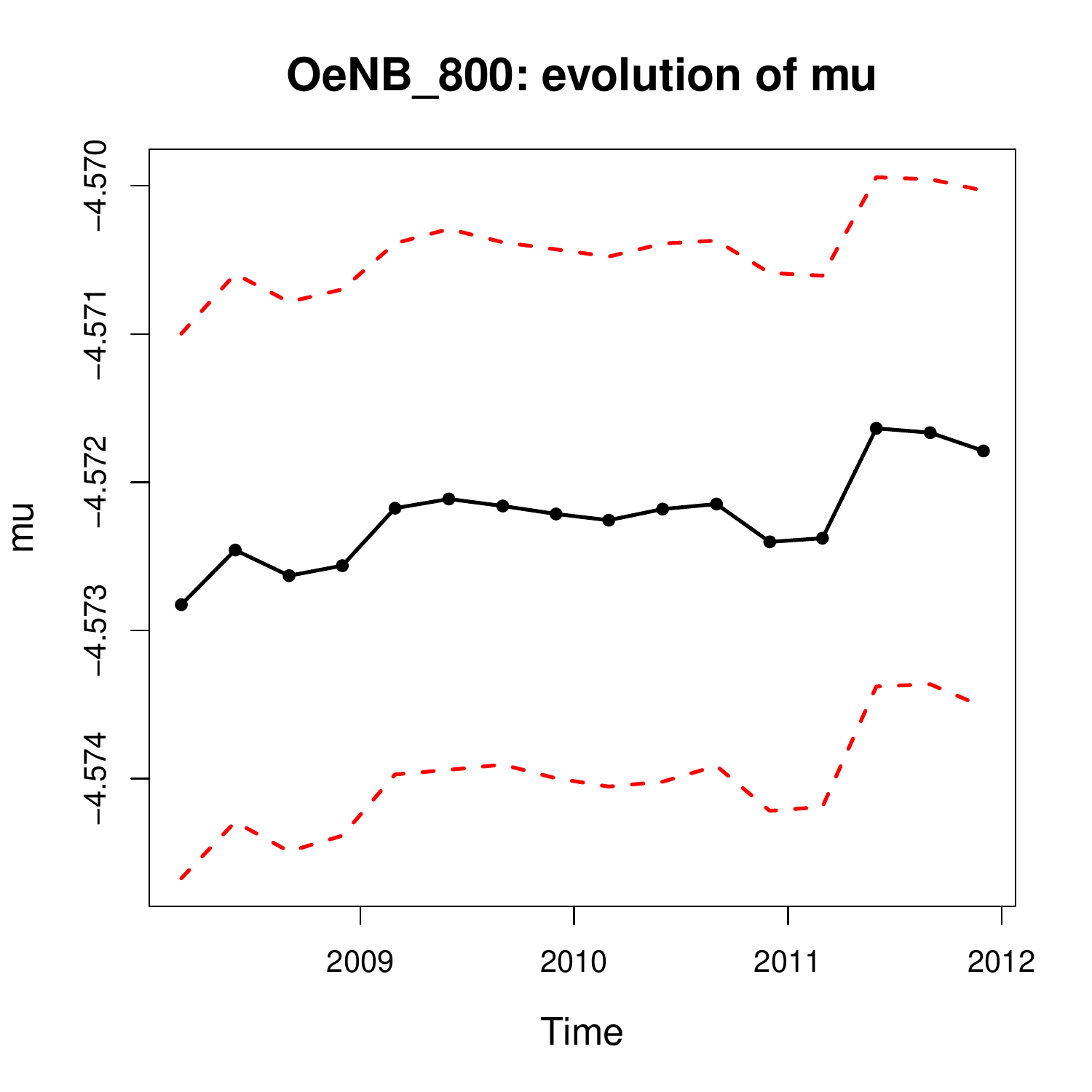}}
		\hspace{0.05\linewidth}
		&
		\subfloat[OeNB 100]{
			\label{fig:AvgMu:b}
			\includegraphics[width=0.45\linewidth]{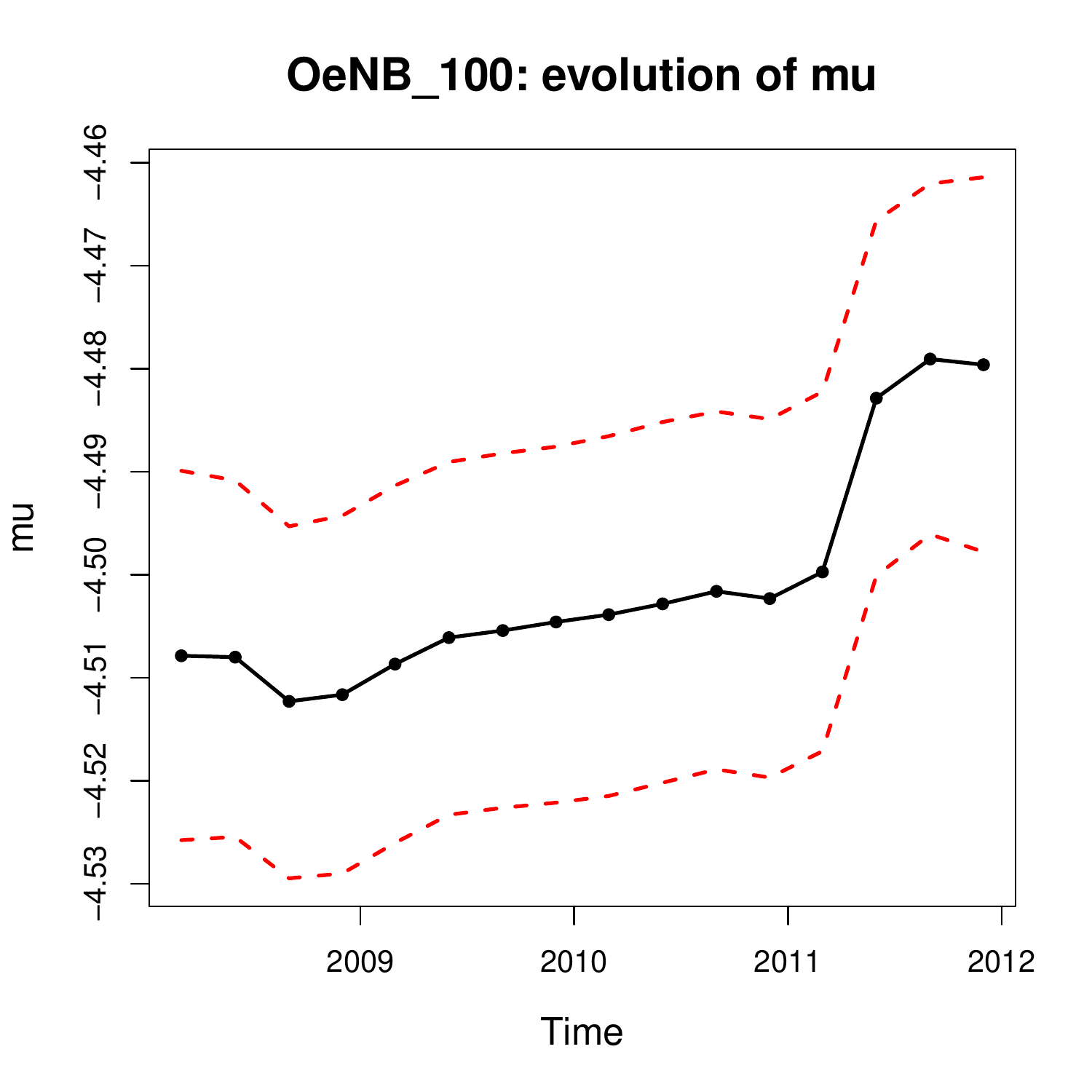}}
	\end{tabular}
	\caption{Evolution of the posterior mean of $\mu_t$ for the full sample (a) and the sample containing only the 100 most relevant banks (b), with $95\%$ credible intervals.}
	\label{fig:AvgMu} 
\end{figure}

First, we study the diversification of the banks which translates to changes in network homogeneity. The drift parameter $\mu_t$, shown in Figure \ref{fig:AvgMu}, exhibits an upward trend for both datasets.
This trend is in both cases more pronounced during the onset of the 2011 sovereign debt crisis. Furthermore, we observe a sharper increase in \texttt{OeNB 100} during this time period. This signals that larger and systemically relevant banks were the ones with a particularly strong reaction to the crisis in comparison to other periods. Interestingly, we do not observe similar behavior during the crisis in 2008.

In the exploratory analysis conducted earlier, we have seen a substantial drop in overall size of exposures in 2008 and almost no such effect in 2011. 
Paradoxically, 2011 is the time when we observe a large upward shift in diversification, while the same effect in 2008 is limited at best. One takeaway from this would be that Austrian banks have perceived the sovereign crisis as a bigger threat than the 2008 crisis stemming from the US housing market. Furthermore, the relative size of this effect is more pronounced in the OeNB 100 sample. This hints at the fact that bigger banks tend to react stronger in the face of adverse conditions by increasing their level of diversification, while less relevant banks tend to keep their exposures less diversified.

Besides the overall development of diversification in the system, we also study the local interaction of banks in the sample. This can be achieved by observing parameters $\theta_i$ which characterize their individual diversification appetite. 

\begin{figure}[!ht]
	\centering
	\begin{tabular}{cc}
		\subfloat[OeNB 800]{
			\label{fig:AvgTheta:a} 
			\includegraphics[width=0.45\linewidth]{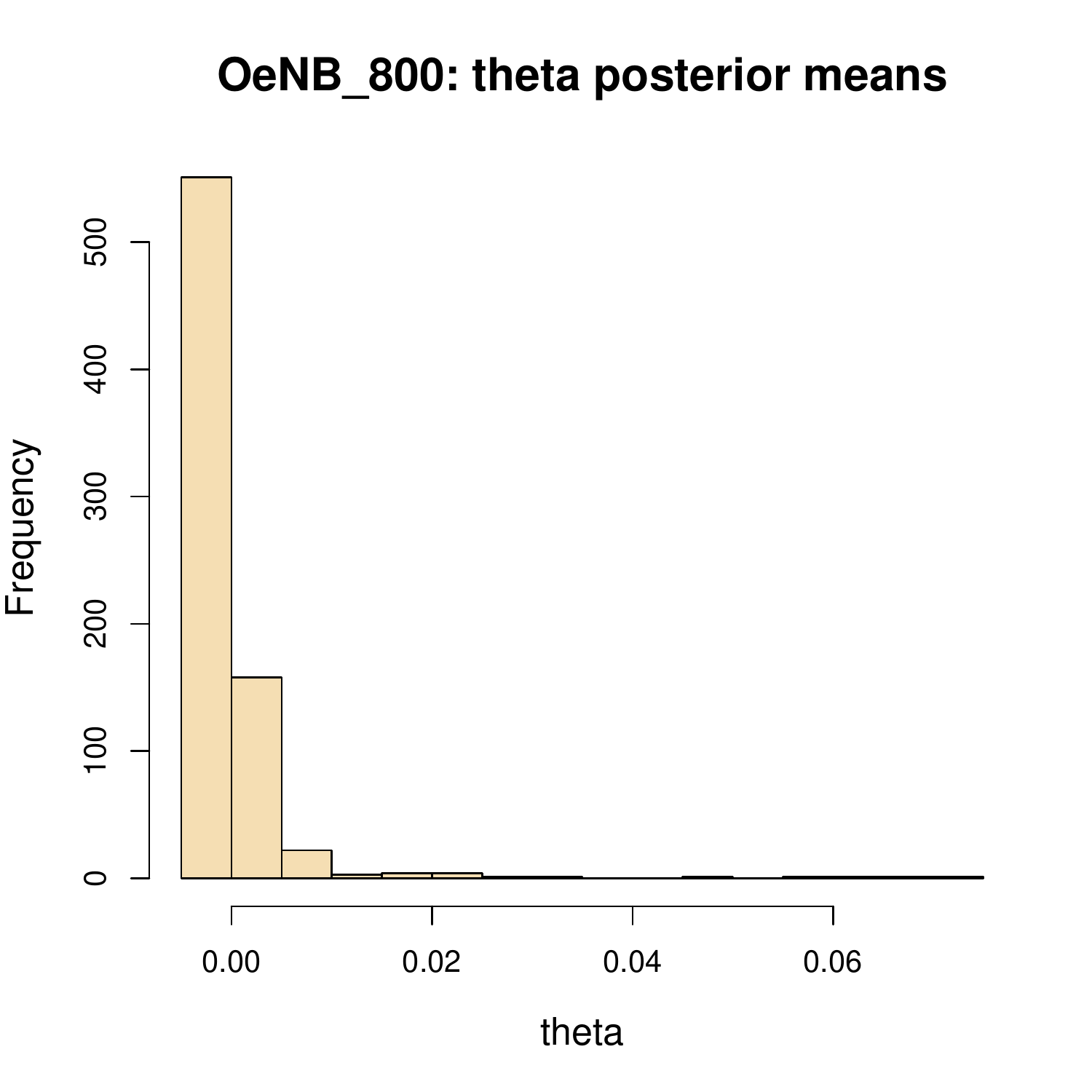}}
		\hspace{0.05\linewidth}
		&
		\subfloat[OeNB 100]{
			\label{fig:AvgTheta:b}
			\includegraphics[width=0.45\linewidth]{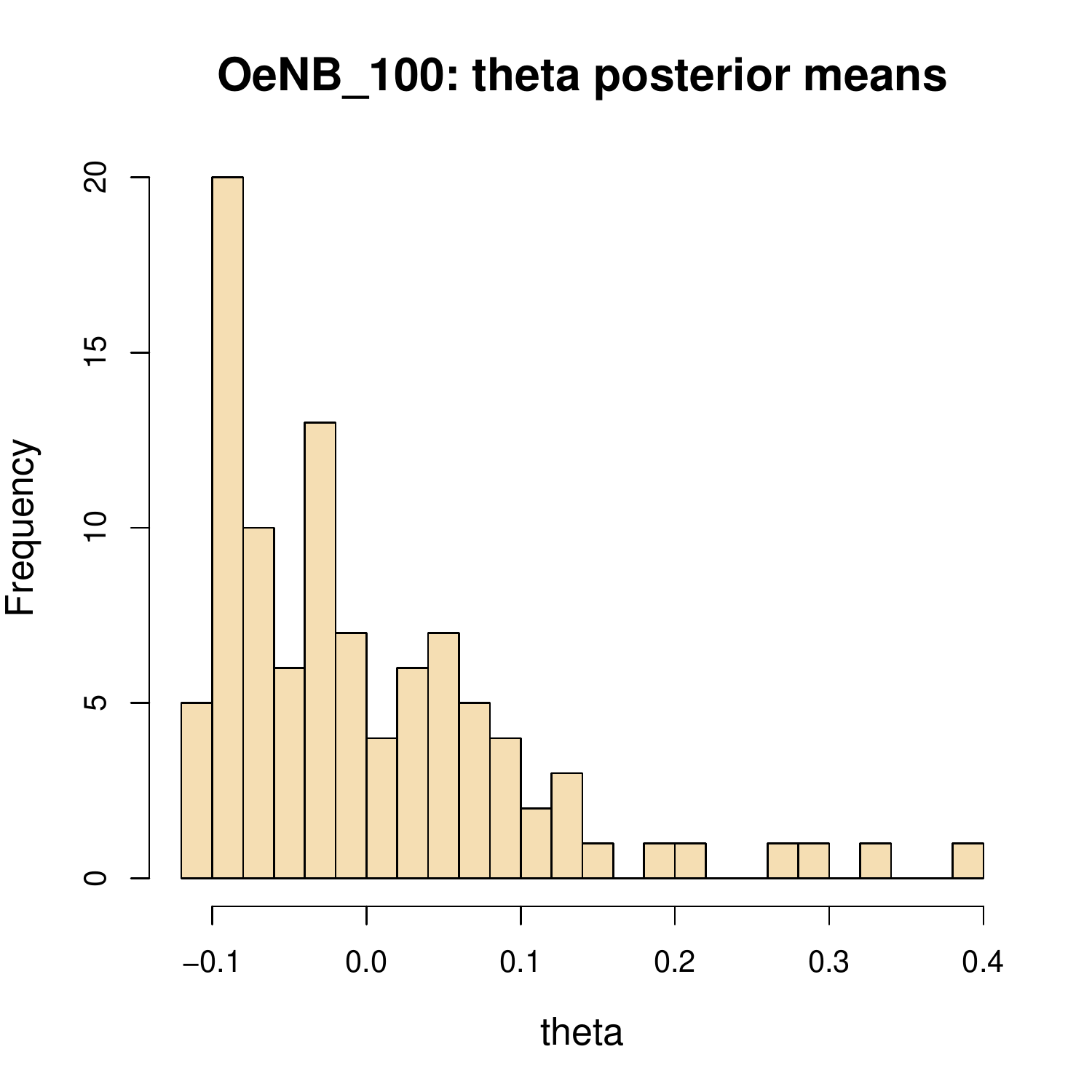}}
	\end{tabular}
	\caption{Posterior distribution of $\boldsymbol{\theta}$ for the full sample (a) and the sample containing only the 100 most relevant banks (b).}
	\label{fig:AvgTheta} 
\end{figure}

\begin{figure}[!ht]
	\centering
	\begin{tabular}{cc}
		\subfloat[OeNB 800]{
			\label{fig:theta_vs_relevance:a} 
			\includegraphics[width=0.45\linewidth]{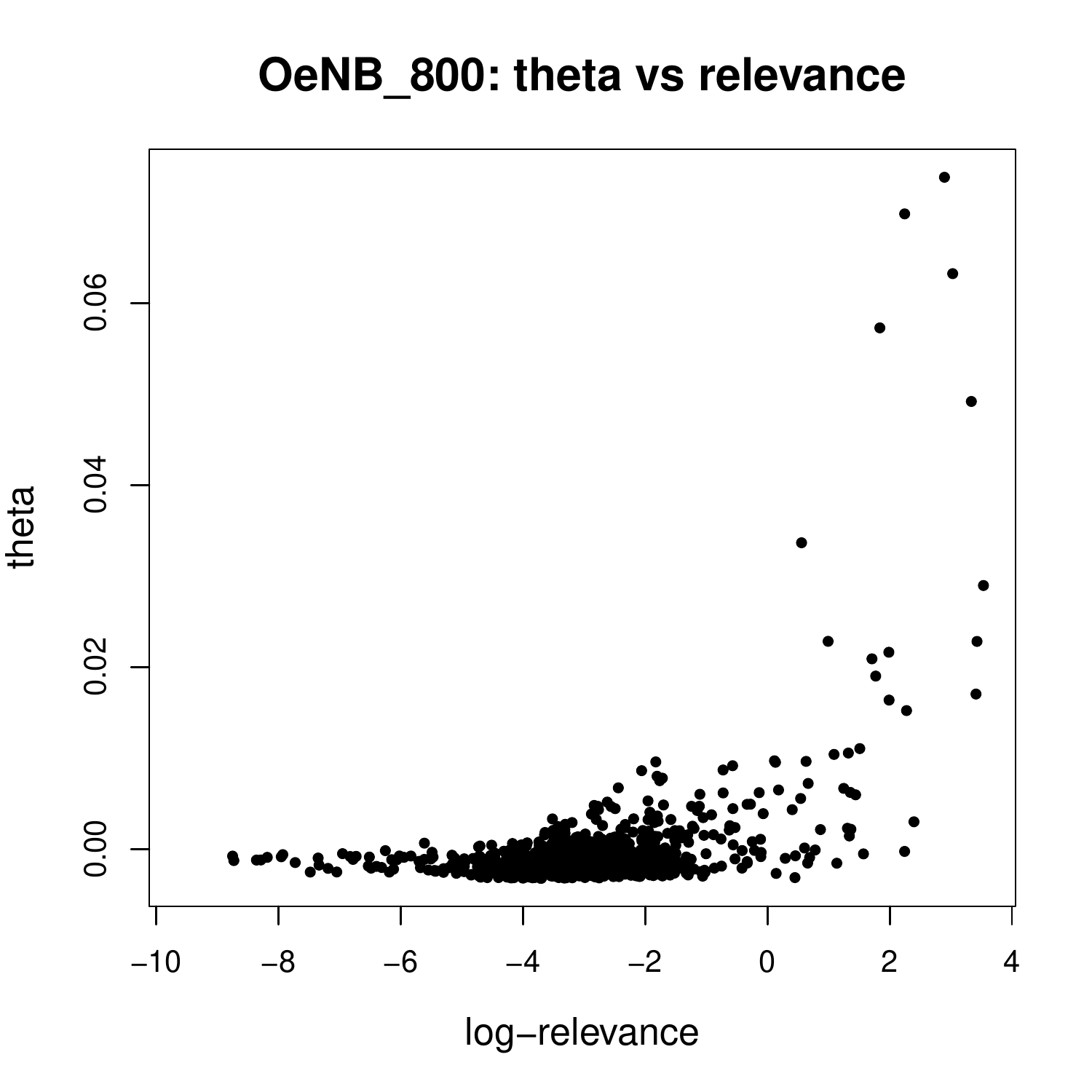}}
		\hspace{0.05\linewidth}
		&
		\subfloat[OeNB 100]{
			\label{fig:theta_vs_relevance:b}
			\includegraphics[width=0.45\linewidth]{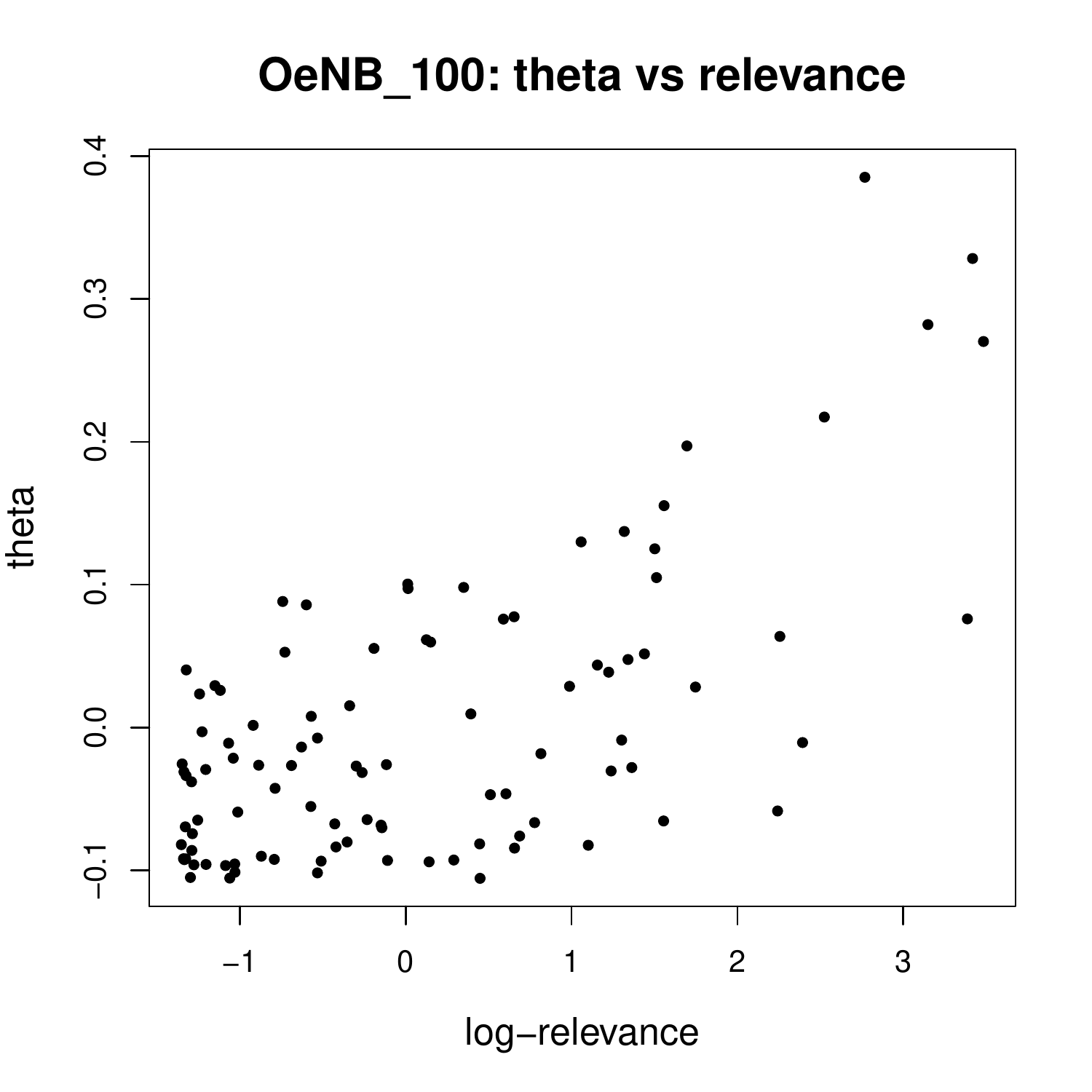}}
	\end{tabular}
	\caption{The banks with higher aggregated relevance tend to also have a higher diversification of exposures in both datasets.}
	\label{fig:theta_vs_relevance} 
\end{figure}

First, we analyze point estimates of these parameters: Figure \ref{fig:AvgTheta} shows the distribution of the posterior means for $\boldsymbol{\theta}$.
For both \texttt{OeNB 800} as well as \texttt{OeNB 100}, the distribution seems to be rather heavy tailed. This translates to a system where the majority of banks exhibits low diversification, but still a fairly large number of banks tends to diversify much more.
In fact, Figure \ref{fig:theta_vs_relevance} highlights that more relevant banks tend to have a more pronounced diversification, whereas small banks do not diversify as much.
This observation further confirms our ideas about a stylized financial network where the disassortative behavior is very common.

\begin{figure}[!ht]
	\centering
	\begin{tabular}{cc}
		\subfloat[OeNB 800]{
			\label{fig:AvgGamma:a} 
			\includegraphics[width=0.45\linewidth]{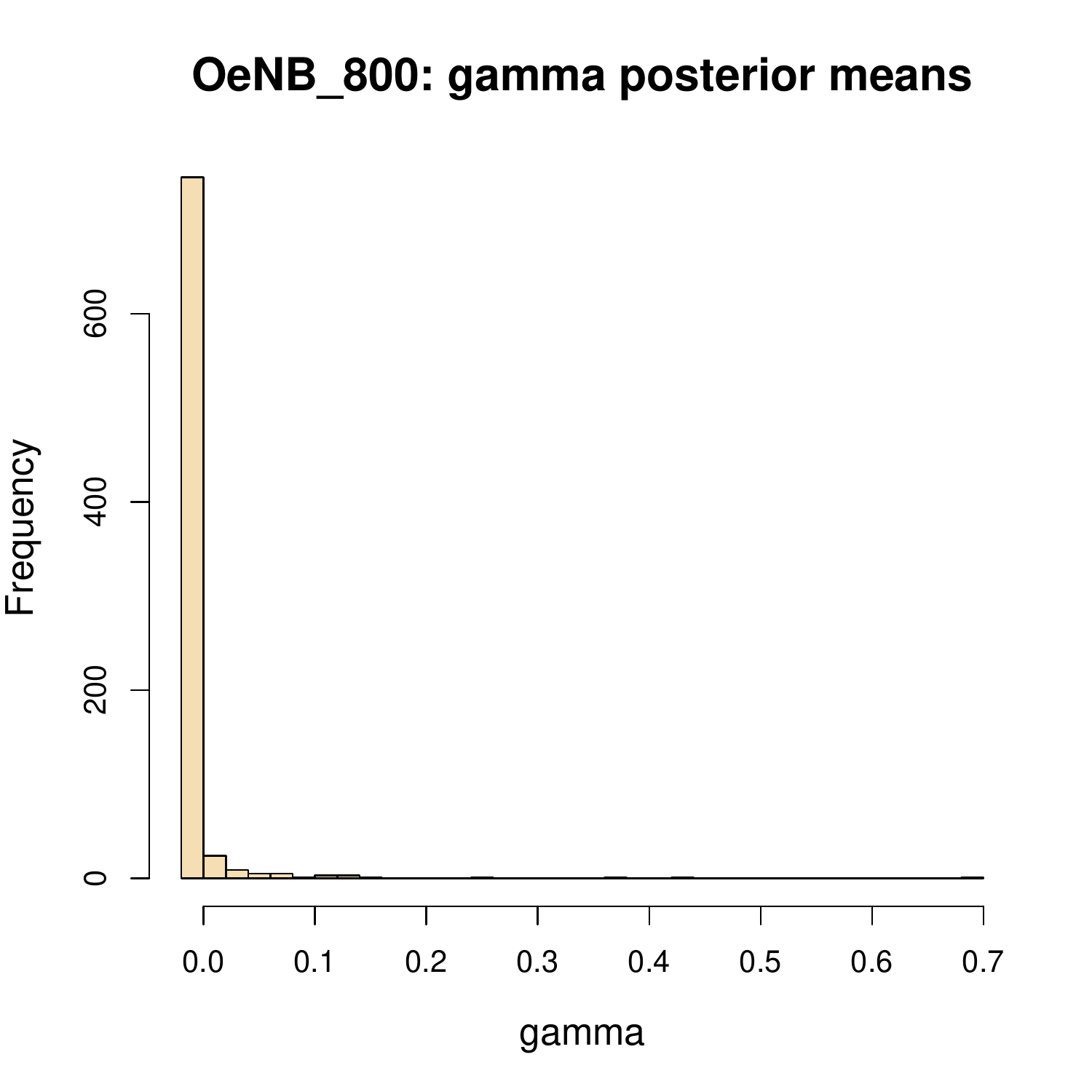}}
		\hspace{0.05\linewidth}
		&
		\subfloat[OeNB 100]{
			\label{fig:AvgGamma:b}
			\includegraphics[width=0.45\linewidth]{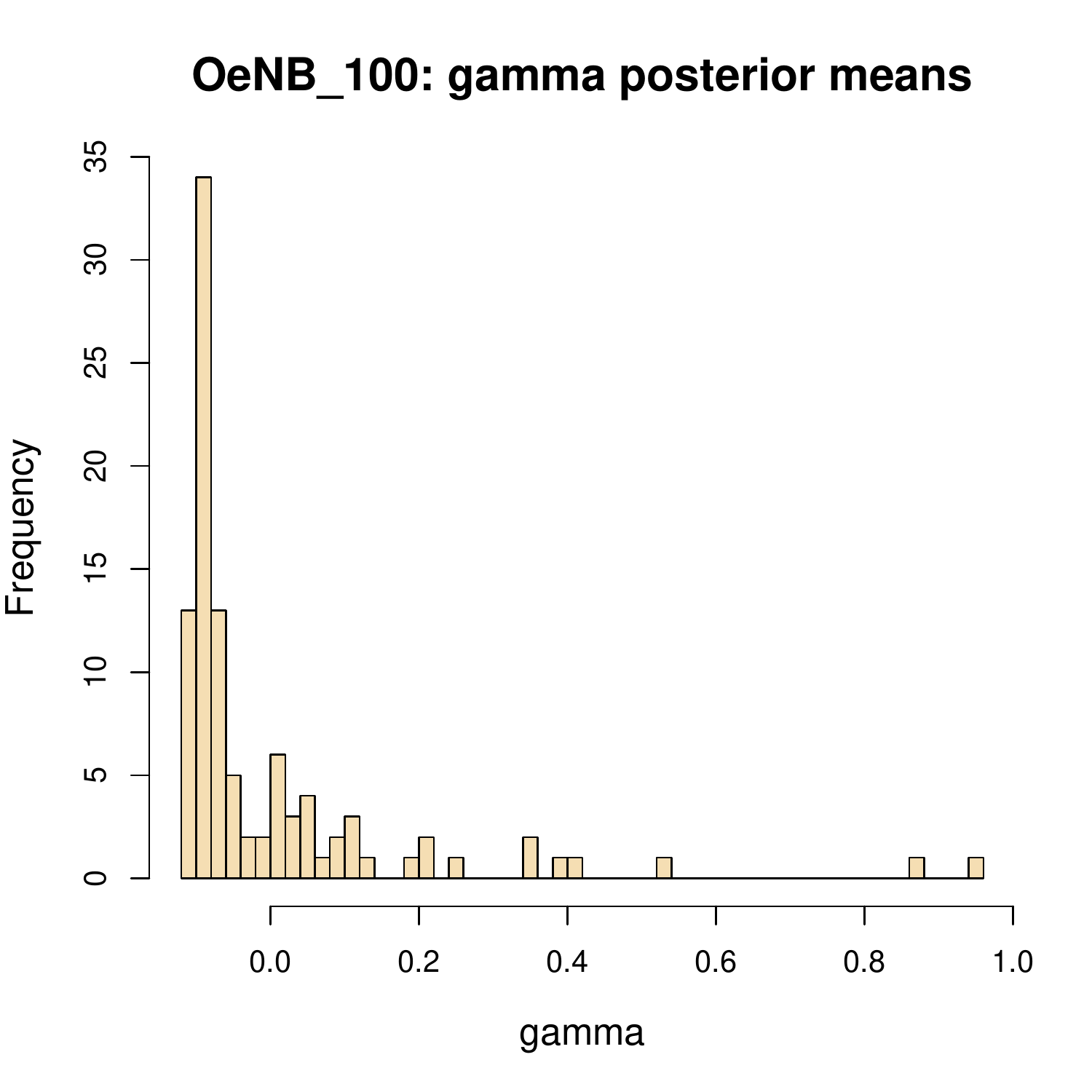}}
	\end{tabular}
	\caption{Posterior distribution of $\boldsymbol{\gamma}$ for the full sample (a) and the sample containing only the 100 most relevant banks (b).}
	\label{fig:AvgGamma} 
\end{figure}

\begin{figure}[!ht]
	\centering
	\begin{tabular}{cc}
		\subfloat[OeNB 800]{
			\label{fig:ThetaVsGamma:a} 
			\includegraphics[width=0.45\linewidth]{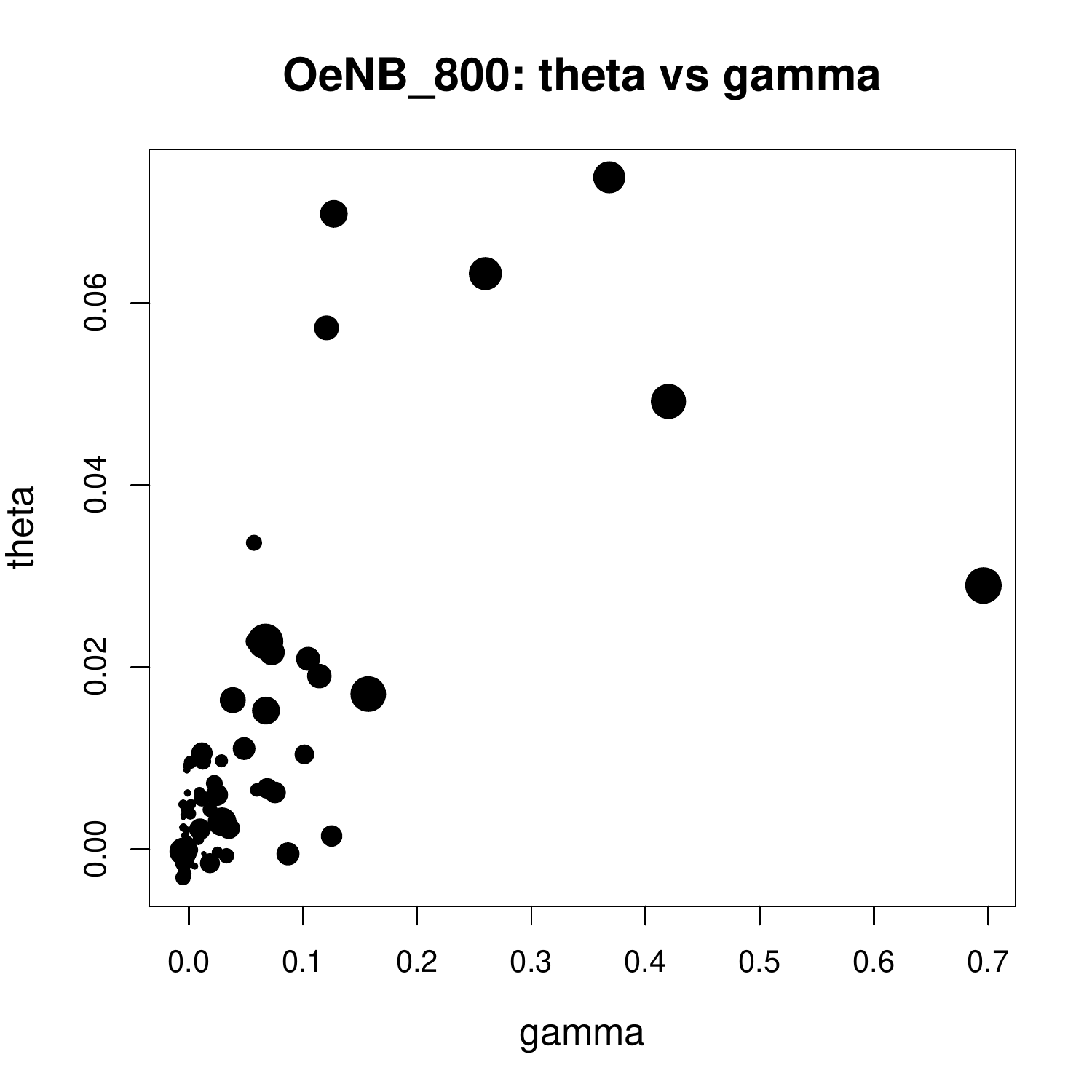}}
		\hspace{0.05\linewidth}
		&
		\subfloat[OeNB 100]{
			\label{fig:ThetaVsGamma:b}
			\includegraphics[width=0.45\linewidth]{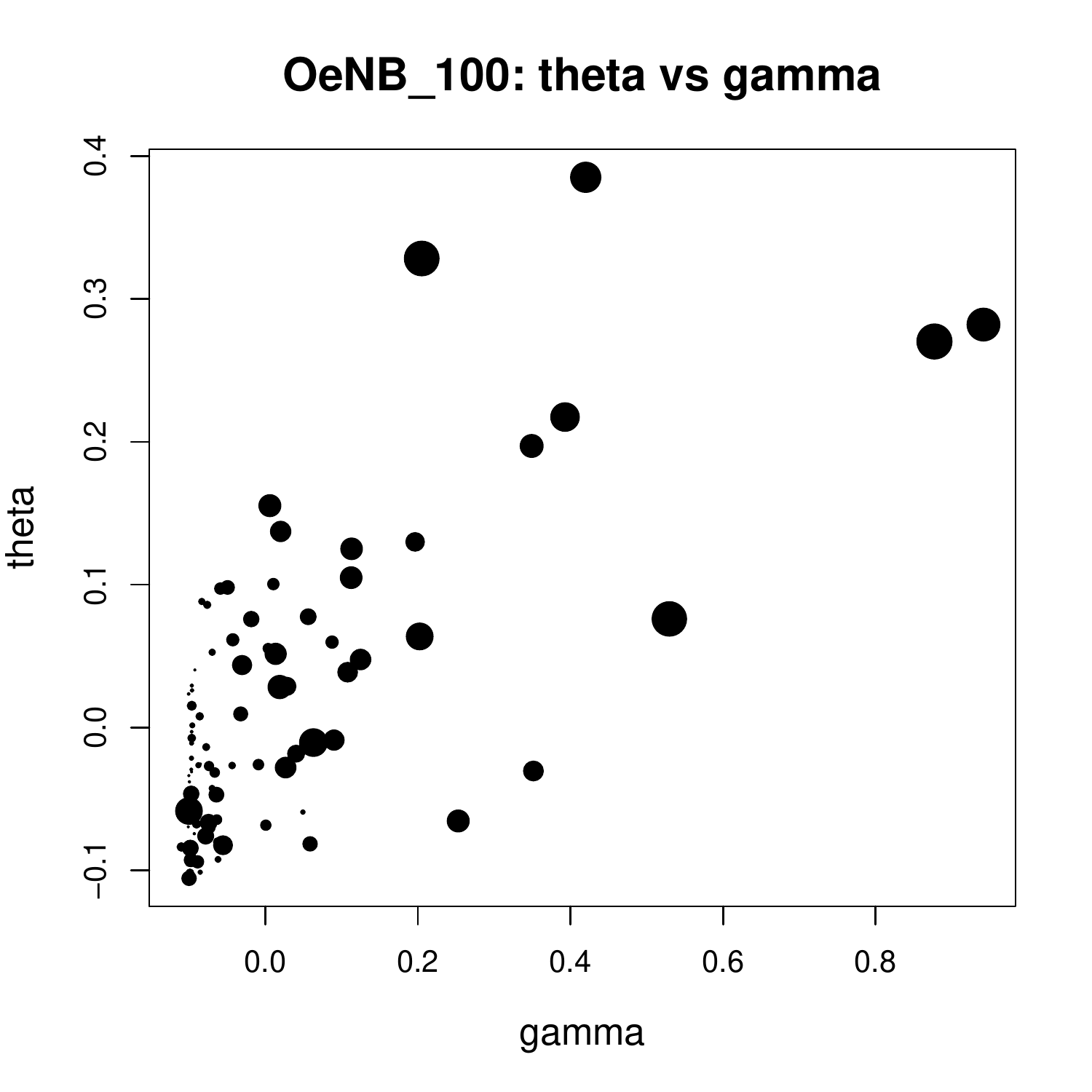}}
	\end{tabular}
	\caption{Posterior distribution of $\gamma$ for the full sample (a) and the sample containing only the 100 most relevant banks (b). In both plots, the size of each circle represents the aggregated relevance of the corresponding bank.}
	\label{fig:ThetaVsGamma} 
\end{figure}

A similarly heavy tailed distribution can be observed regarding the attractiveness parameter $\boldsymbol{\gamma}$ (see Figure \ref{fig:AvgGamma} for the distribution of the point estimates, where the heavy right tail is apparent).
In addition, Figure \ref{fig:ThetaVsGamma} shows that, generally, $\boldsymbol{\theta}$ and $\boldsymbol{\gamma}$ are closely related in both datasets.
This figure highlights that larger banks tend to be more diversified and more attractive simultaneously, and, vice versa, small banks often play a role in the periphery of the network as offsprings of a larger bank. A similar observation of heavy-tailedness in degree distribution has also been reported by \cite{Boss2004}.

\begin{figure}[!ht]
	\centering
	\begin{tabular}{cc}
		\subfloat[OeNB 800]{
			\label{fig:variance:a} 
			\includegraphics[width=0.45\linewidth]{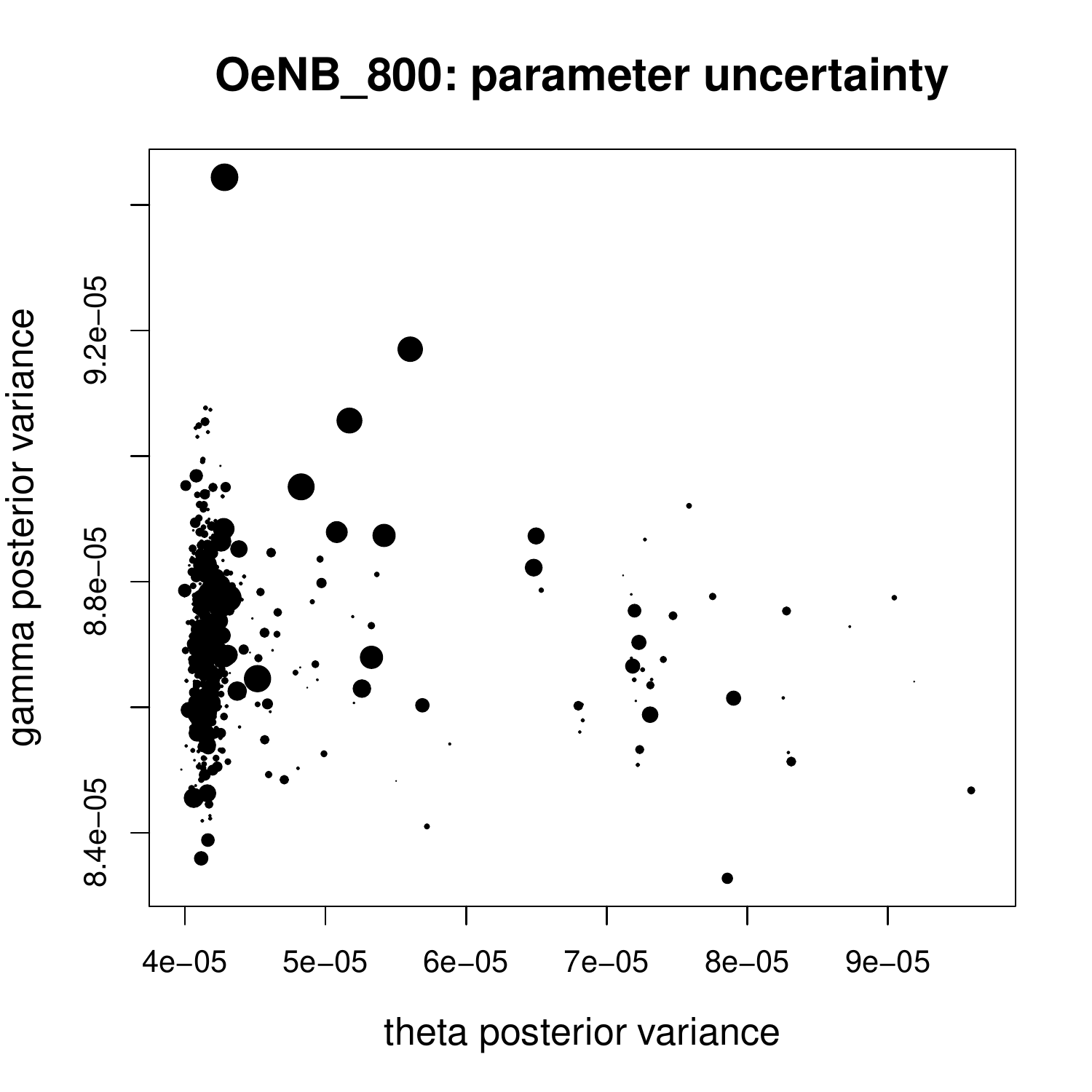}}
		\hspace{0.05\linewidth}
		&
		\subfloat[OeNB 100]{
			\label{fig:variance:b}
			\includegraphics[width=0.45\linewidth]{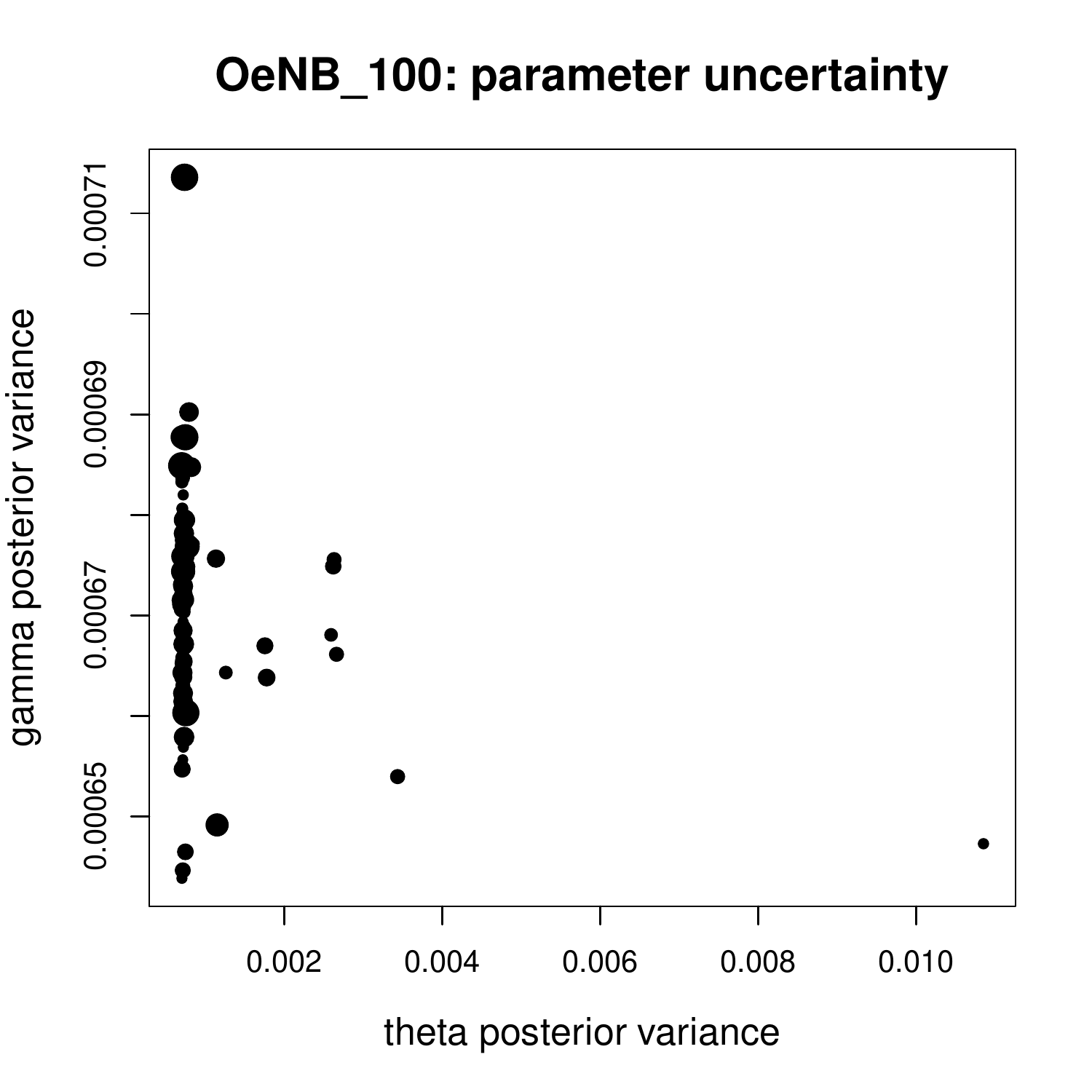}}
	\end{tabular}
	\caption{Posterior variances of $\boldsymbol{\theta}$ and $\boldsymbol{\gamma}$ for the full sample (a) and the sample containing only the 100 most relevant banks (b). In both plots, the size of each circle represents the aggregated relevance of the corresponding bank.}
	\label{fig:variance} 
\end{figure}

As concerns the uncertainty around the point estimates, Figure \ref{fig:variance} compares the posterior variances for all of the $\boldsymbol{\theta}$ with those of the $\mu_s$
We note that there seems to be no explicit pattern and no apparent relation with the relevance of the corresponding banks.
We point out, however, that the two plots are on two different scales on both axes, which is expected since much more data is available for inference in the \texttt{OeNB 800} dataset, hence yielding more reliable estimates.

\begin{figure}[!ht]
	\centering
	\begin{tabular}{cc}
		\subfloat[$1/\tau_\eta$, \texttt{OeNB 800}]{
			\label{fig:TauDensity:a} 
			\includegraphics[width=0.35\linewidth, page=1]{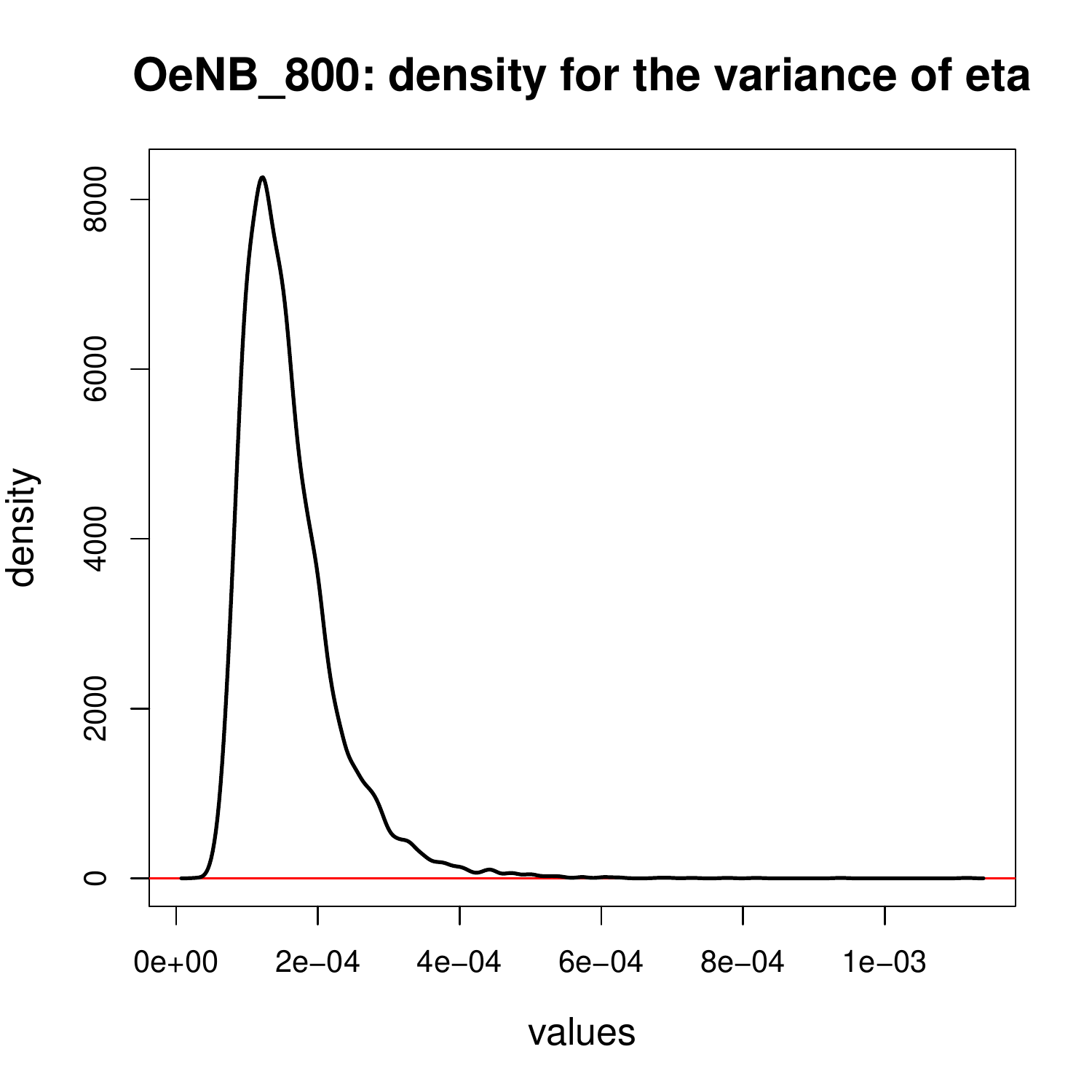}}
		\hspace{0.05\linewidth}
		&
		\subfloat[$1/\tau_\eta$, \texttt{OeNB 100}]{
			\label{fig:TauDensity:b}
			\includegraphics[width=0.35\linewidth, page=1]{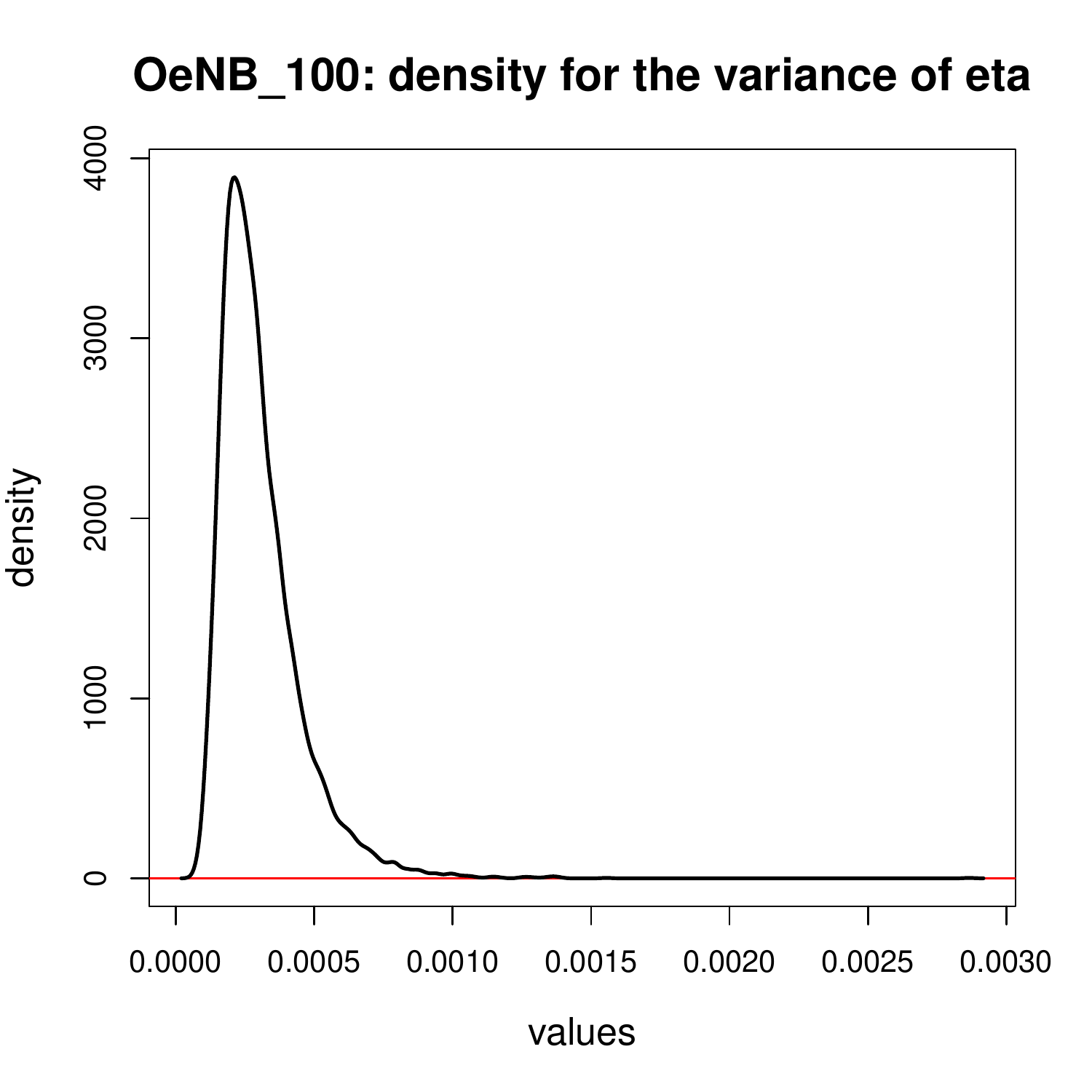}}\\
		\subfloat[$1/\tau_\theta$, \texttt{OeNB 800}]{
			\label{fig:TauDensity:c} 
			\includegraphics[width=0.35\linewidth, page=2]{OeNB_800_tau_density}}
		\hspace{0.05\linewidth}
		&
		\subfloat[$1/\tau_\theta$, \texttt{OeNB 100}]{
			\label{fig:TauDensity:d}
			\includegraphics[width=0.35\linewidth, page=2]{OeNB_100_tau_density}}\\
		\subfloat[$1/\tau_\gamma$, \texttt{OeNB 800}]{
			\label{fig:TauDensity:e} 
			\includegraphics[width=0.35\linewidth, page=3]{OeNB_800_tau_density}}
		\hspace{0.05\linewidth}
		&
		\subfloat[$1/\tau_\gamma$, \texttt{OeNB 100}]{
			\label{fig:TauDensity:f}
			\includegraphics[width=0.35\linewidth, page=3]{OeNB_100_tau_density}}\\						
	\end{tabular}
	\caption{Posterior distribution of variance parameters $1/\tau_\eta, 1/\tau_\theta$ and $1/\tau_\gamma$ for \texttt{OeNB 800} (left) and \texttt{OeNB 100} (right) datasets.}
	\label{fig:TauDensity} 
\end{figure}

Finally, we also show the posterior densities for the variance parameters $1/\tau_\eta, 1/\tau_\theta$ and $1/\tau_\gamma$ in Figure \ref{fig:TauDensity}.
For both datasets, these plots suggest that the drift parameter is rather stable over time, and that the diversification and attractiveness are not particularly diverse across banks, overall.

\section{Conclusion}
This paper contibutes to the networks literature by proposing a brand new framework to model the evolution of dynamic weighted networks, and to capture systematic parts of their development. 
Our application to the Austrian interbank market gives a new perspective on the recent crises and demonstrates how our model can be used in as a means to measure exposure diversification and, hence, one aspect of systemic risk. Differently from \cite{friel2016interlocking}, our measure is not affected by banks entering or leaving the system, since our dataset only contains banks which are active throughout the whole period. 
In our analysis we have shown that the Austrian market exhibited a sustained increase in banks' diversification, possibly as a reaction to the 2008 financial crisis.
In particular, differently from a descriptive analysis, our model captured a distinct upward dynamic in network homogeneity as a response to the sovereign debt crisis of 2011.
These findings may be of a particular use to regulators and central banks to assess and design future policy.

Our results also showed that the roles played by the different banks can be vastly different, particularly in the context of exposure diversification.
Our findings emphasize that larger banks, which are generally more susceptible to systemic risk, tend to use more conservative strategies and to spread out evenly their credit risks.

One limitation of our modeling framework is that it only focuses on the relative exposures, hence discarding the real magnitudes of the claims.
Future extensions of this work may consider a joint modeling of the exposure values and how they are diversified among neighbors.

Another possible extension of our framework would include a more sophisticated prior structure on the model parameters.
For example, one may define a clustering problem on the banks, where different clusters are characterized by different network homogeneity drifts $\boldsymbol{\mu}$.

Finally, we would like to remark that the Dirichlet likelihood specification is not the only possible one.
Besides, the Dirichlet distribution is known to exhibit very little flexibility, since, when the variance is large, it tends to assign most of the probability density to the highest entropy configurations.
This does not necessarily reflect the features exhibited by the data.
However, we argue that in our application the Dirichlet assumption is very reasonable, and, more importantly, it provides a convenient framework with a straightforward interpretation of the model parameters.

\bibliographystyle{jf}
\bibliography{references}{}	

\begin{appendices}
\section{Data Transformation}\label{app:data_transormation}
The source data from the Austrian National Bank is in the form of four variables: a timestamp, an ID of a lender bank, an ID of a borrower, and the \textit{relative} exposure from one towards the other. We use the term \textit{relative} since the largest exposure in each time period is assumed to be of size 1, and all other exposures in that time period are scaled accordingly to keep their relative size unchanged. As a result, in each time-period, all exposures are located in a $\left(0,1\right]$ interval with the highest exposure attaining a value of 1. Formally, making use of Definition \ref{def:TrueExposures} for true exposures, the observable data in our sample can be viewed as a dynamic adjacency matrix $D$:

\begin{definition}
\label{def:DataExposures}
A \textit{sequence of observable exposures} $\mathcal{D}=\{{D}_t\}_{t\in \mathcal{T}}$ on the set of nodes $\mathcal{V}$ over the timespan $\mathcal{T}$ is defined as follows:

\begin{equation}
d_{ij}^{(t)}\stackrel{def}{=}\frac{e_{ij}^{(t)}}{\max_{\substack{k,l}} e_{kl}^{(t)}} \ \ \ \forall i,j,k,l \in \mathcal{V}, \forall t \in \mathcal{T}
\end{equation}
\end{definition}

It is not possible to make inter-temporal analysis of changes in exposures while working directly with sequence $\mathcal{D}$, because every exposure is scaled against the highest exposure in its time period. In order to circumvent this issue and obtain information which is comparable in time, we have devised the following procedure.

We make an assumption about the stability of the Austrian market. Namely, when looking at the change of a particular edge value between two consecutive periods from $d_{ij}^{(t)}$ to $d_{ij}^{(t+1)}$, the ratio $\frac{d_{ij}^{(t)}}{d_{ij}^{(t+1)}}$ with highest likelihood of occurrence in the sample corresponds to banks keeping the absolute value of their exposures unchanged. Indeed, after examining this ratio in all consecutive periods, we observe that the most frequent value is situated in the middle of the sample and is always a clear outlier in terms of likelihood of occurrence.\footnote{In most cases, this value is around 1 which suggests that the largest exposure in the network is mostly stable. An exception arises between dates 2 and 3 which correspond to the second and third quarter of 2008. As this is the exact time of the height of US subprime mortgage crisis, we believe that the ``big players'' in our dataset have been influenced by these events, resulting in the change of their exposures and subsequent substantial rescaling of the whole system. According to our methodology, the largest exposure in the network has dropped to almost one third of its value in the span of two quarters, but it returns gradually back to its former level eventually.}

It's straightforward to rescale the whole dataset using this procedure. Despite the fact that we still cannot observe the actual levels of exposures between banks in our sample, we are now able to compare them inter-temporally which is an extremely useful property. We will be referring to such rescaled dataset as the \textit{sequence of absolute exposures} and denote it by $\mathcal{X}$.

We only have a single use for $\mathcal{X}$, namely to create a subsample of ``core banks''. In fact, selecting a portion of banks which can be deemed important allows us to see how the implications of our model are affected by the banks' size. In order to do so, we introduce the bank's \textit{relevance}:
\begin{definition}
	A relevance of bank $i$ in time period $t$ is defined as:
	\begin{equation}
	r_i^{(t)} = \sum_{k\in\mathcal{N}} x_{ik}^{(t)} + \sum_{k\in\mathcal{N}} x_{ki}^{(t)}.
	\end{equation}
\end{definition}
In other words, we define relevance simply as the bank's overall sum of its interbank assets and liabilities.

With a clear measure of systemic importance, we can now select a subsample of banks with the highest \textit{aggregated relevance} $r_i = \sum_{t=1}^{T} r_i^{(t)}$. This allows us to focus on the interactions of systemically important banks and observe whether there are some unique patterns. 
We use the aggregated relevance measure to create a new smaller dataset consisting of the 100 most systemically relevant institutions and the connections between them. 
We shall refer to the full dataset and the reduced dataset as \texttt{OeNB 800} and \texttt{OeNB 100}, respectively\footnote{Validity of the \texttt{OeNB 100} subset can be justified further by examining the overall exposure of top 100 institutions. It turns out that the 100 most systemically relevant banks account for more than 95\% of all edge weights in any given time frame. In other words, the 100 most systemically relevant banks are the ones responsible for the vast majority of all exposures within the system, which makes their closer examination interesting}.

Lastly, for modeling purposes, we define the sequence of relative exposures as follows:

\begin{definition}
	\label{def:RelativeExposures}
	A \textit{sequence of relative exposures} $\mathcal{Y}=\{{Y}_t\}_{t\in \mathcal{T}}$ on the set of nodes $\mathcal{V}$ over the timespan $\mathcal{T}$ has elements defined as follows:
	\begin{equation}
	y_{ij}^{(t)}\stackrel{def}{=}\frac{x_{ij}^{(t)}}{\sum_{k=1}^{N} x_{ik}^{(t)}} \ \ \ \forall i,j \in \mathcal{V}, \forall t \in \mathcal{T}
	\end{equation}
\end{definition}

To summarize, there are four different types of dynamic adjacency matrices used in our paper: $\mathcal{E}$ corresponds to the true data with the actual connection values which we do not observe, $\mathcal{D}$ represents the scaled data where edge weights are normalized with respect to the highest value in \textit{each} period, $\mathcal{X}$ contains the scaled data where all edge weights are normalized with respect to the highest value in the \textit{first} period, and $\mathcal{Y}$ contains the relative exposures of banks derived from $\mathcal{X}$ which makes them comparable intertemporally. We are working mainly with $\mathcal{Y}$ as it contains the most useful intormation we can get from the data.
\end{appendices}

\end{document}